\documentclass[12pt]{article}
\usepackage{amssymb}
\usepackage{amsmath}
\usepackage{color}
\usepackage{extarrows}
\usepackage[section]{placeins}
\usepackage[figuresright]{rotating}
\usepackage{multirow}
\usepackage{caption}
\usepackage{tikz}
\usepackage{graphics}
\usepackage{cite}

\begin{document}



\def\a{\alpha}
\def\b{\beta}
\def\d{\delta}
\def\e{\epsilon}
\def\g{\gamma}
\def\k{\kappa}
\def\l{\lambda}
\def\o{\omega}
\def\p{\wp}
\def\r{\rho}
\def\t{\tau}
\def\s{\sigma}
\def\z{\zeta}
\def\x{\xi}
\def\V={{{\bf\rm{V}}}}
 \def\A{{\cal{A}}}
 \def\B{{\cal{B}}}
 \def\C{{\cal{C}}}
 \def\D{{\cal{D}}}
\def\D{\Delta}
\def\G{\Gamma}
\def\K{{\cal{K}}}
\def\O{\Omega}
\def\R{\bar{R}}
\def\T{{\cal{T}}}
\def\L{\Lambda}

\def\R{\overline{R}}

\def\bee{\begin{equation}}
\def\eee{\end{equation}}
\def\be{\begin{eqnarray}}
\def\ee{\end{eqnarray}}
\def\ba{\begin{array}}
\def\ea{\end{array}}
\def\no{\nonumber}
\def\le{\langle}
\def\re{\rangle}
\def\lt{\left}
\def\rt{\right}
\def\mc{\mathcal}
\def\sc{\scriptscriptstyle}

\newtheorem{Theorem}{Theorem}
\newtheorem{Definition}{Definition}
\newtheorem{Proposition}{Proposition}
\newtheorem{Lemma}{Lemma}
\newtheorem{Corollary}{Corollary}
\newcommand{\proof}[1]{{\bf Proof. }
        #1\begin{flushright}$\Box$\end{flushright}}
\renewcommand{\theequation}{\thesection.\arabic{equation}}
\renewcommand{\thetable}{\thesection.\arabic{table}}
\renewcommand{\thefigure}{\thesection.\arabic{figure}}

\begin{titlepage}
\begin{center}

    \LARGE $T$-$Q$ relations for the integrable two-species asymmetric simple exclusion process with open boundaries\\
\vspace{1in}
\large Xin Zhang${}^{a,1}$, Fakai Wen${}^{b,2}$, Jan de Gier${}^{a,3}$\\
\vspace{1in}
\normalsize
{\it ${}^a$ ARC Centre of Excellence for Mathematical and Statistical Frontiers (ACEMS), School of Mathematics and Statistics, The University
of Melbourne, VIC 3010, Australia\\
 ${}^b$ State Key Laboratory of Magnetic Resonance and Atomic and Molecular Physics, Wuhan Institute of Physics and Mathematics, Chinese Academy of Sciences,
Wuhan 430071, China}

\vspace{0.3in}
\normalsize
{E-mail:~ ${}^1$zhang.x1@unimelb.edu.au, ${}^2$fakaiwen@wipm.ac.cn,
 ${}^3$jdgier@unimelb.edu.au}
\end{center}

\vspace{.5in}
\begin{abstract}
We study the integrable two-species asymmetric simple exclusion process (ASEP) for two inequivalent types of open, non particle conserving boundary conditions. Employing the nested off-diagonal Bethe ansatz method, we construct for each case the corresponding homogeneous $T$-$Q$ relations and obtain the Bethe ansatz equations. Numerical checks for small system sizes show completeness for some Bethe ansatz equations, and partial completeness for others.

\vspace{1truecm}

\end{abstract}

\end{titlepage}
\section*{Introduction}
The asymmetric simple exclusion process (ASEP) \cite{spitzer1970,derrida1998,liggett2012} is one of the simplest examples to describe the asymmetric diffusion of hard-core particles with anisotropic hopping rates.
The ASEP is one of the best studied models in non-equilibrium statistical mechanics \cite{derrida1993exact,schutz1993} and plays important roles in a variety fields such as biology\cite{chowdhury2005}, networks \cite{neri2011} and traffic modeling \cite{karimipour1999}.

The ASEP is an integrable system \cite{schutz2001,golinelli2006} which is closely related to the XXZ spin chain \cite{sandow1994,essler1996}. Both periodic and open boundary conditions have been extensively studied \cite{derrida1993exact,sasamoto1999,jan2005,jan2011,fa2015spectrum}. The most well known methods to solve the eigenvalue problem for the open ASEP are the Bethe ansatz \cite{derrida1999,jan2006,jan2008,cantini2008,simon2009,prolhac2010,crampe2010,crampe2015} and the use of matrix product states \cite{derrida1993,prolhac2009}.

The multi-species ASEP (m-ASEP) is also exactly solvable with periodic boundaries \cite{ferrari2007,evans2009} and a variety of open boundaries \cite{cantini2008,cantini2016,Crampe,crampe2016integrable}. While a number of results is known for the exact solution of m-ASEP with periodic boundary conditions, not much is known for open boundary conditions. For m-ASEP with non-diagonal open boundary conditions, particle conservation is completely or partially violated. As a result, an obvious reference state or pseudo-vaccuum is absent and conventional Bethe ansatz methods may fail.

One possible approach to resolve the nontrivial problem of diagonalising the m-ASEP generator is the modified Bethe ansatz \cite{cao2003,nepomechie2003,Rafael2003,Rafael2004}, which has been used to solve the eigenvalue problem of ASEP with open boundary conditions \cite{jan2005}. Another effective method is the off-diagonal Bethe ansatz (ODBA) \cite{cao2013,wang-book}. Several integrable models with nontrivial boundary conditions and high rank were solved by ODBA and the further nested ODBA  \cite{cao2013152,cao2013off,li2014exact,zhang2014exact,cao2014su(n),hao2014}. In 2016, the exact solution of $SU(3)$ XXZ spin chain was given by the nested ODBA \cite{li2016}. This work directly inspired us to solve the eigenvalue problem of two-species ASEP with open boundaries.

The paper is organized as follows. In Section \ref{2-ASEP} we introduce a two-species ASEP with certain non-diagonal open boundary conditions. The integrability of the system and the corresponding transfer matrix are shown in this section. In Section \ref{nested bethe ansatz} we give the construction of a set of fused transfer matrices from which we derive some useful operator identities that are indispensable for the construction of the $T$-$Q$ relation. With the help of these operator identities we propose several homogeneous $T$-$Q$ relations in Section \ref{T-Q relation} and perform some numerical checks for our result. In Section \ref{another-K} we focus on a second set of integrable open boundaries for the two-species ASEP and its exact solution. We discuss extensions to integrable multi-species ASEPs in Section \ref{integrable-mASEP}. Details related to the operator identities are provided in Appendices \ref{A.1} and \ref{A.2}.

\section{An integrable two-species ASEP with open boundaries}\label{2-ASEP}

\subsection{A two-species ASEP with open boundaries}\label{2-ASEP introduction}

\begin{figure}[htp]
  \centering
  \includegraphics[width=10cm]{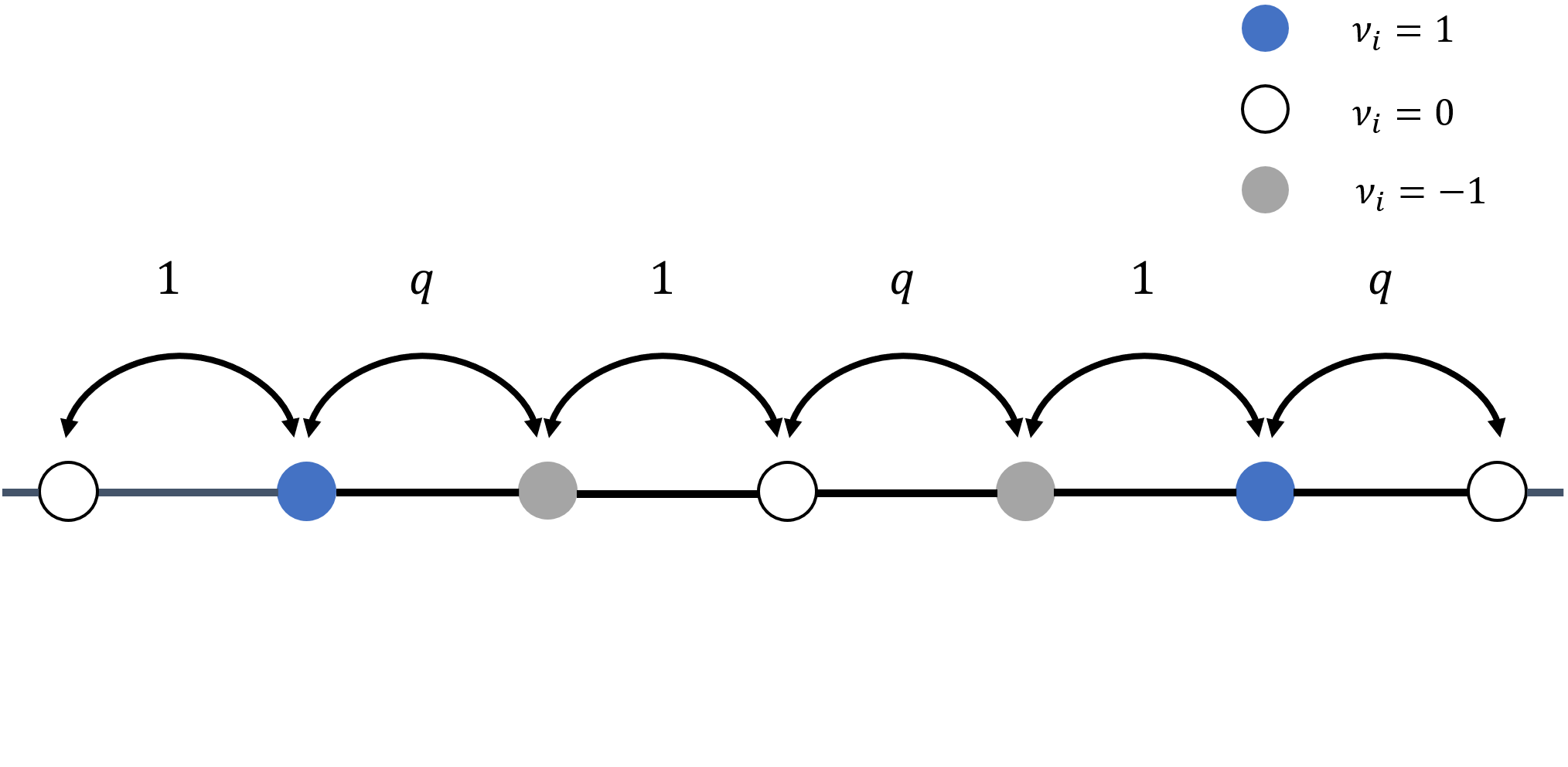}\\
  \caption{The bulk transition rates.}\label{figure-bulk-rate}
\end{figure}

\begin{figure}[htp]
  \centering
  \includegraphics[width=8cm]{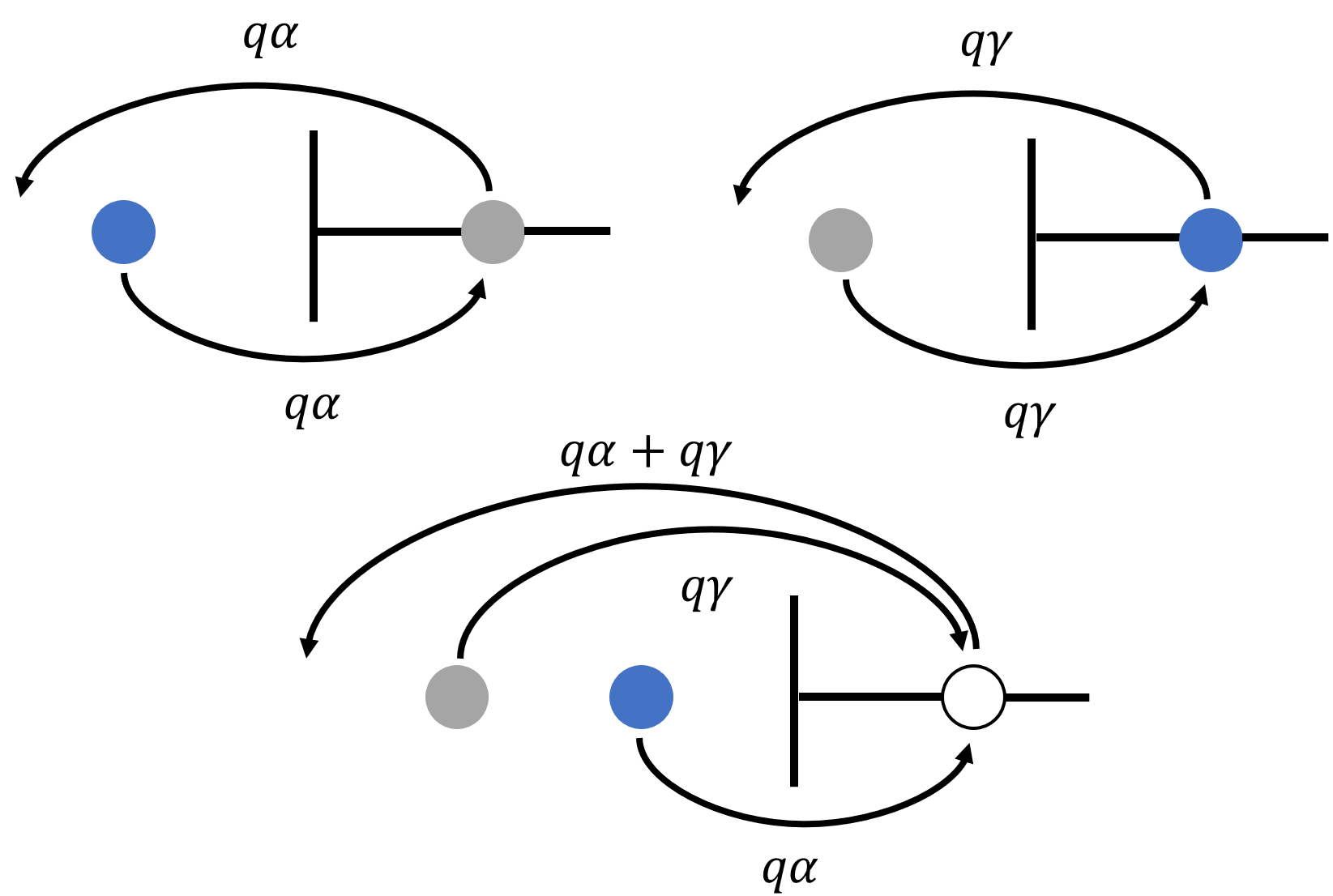}\\
  \caption{The transition rates at left boundary.}\label{figure-left-rate}
\end{figure}

\begin{figure}[htp]
  \centering
  \includegraphics[width=8cm]{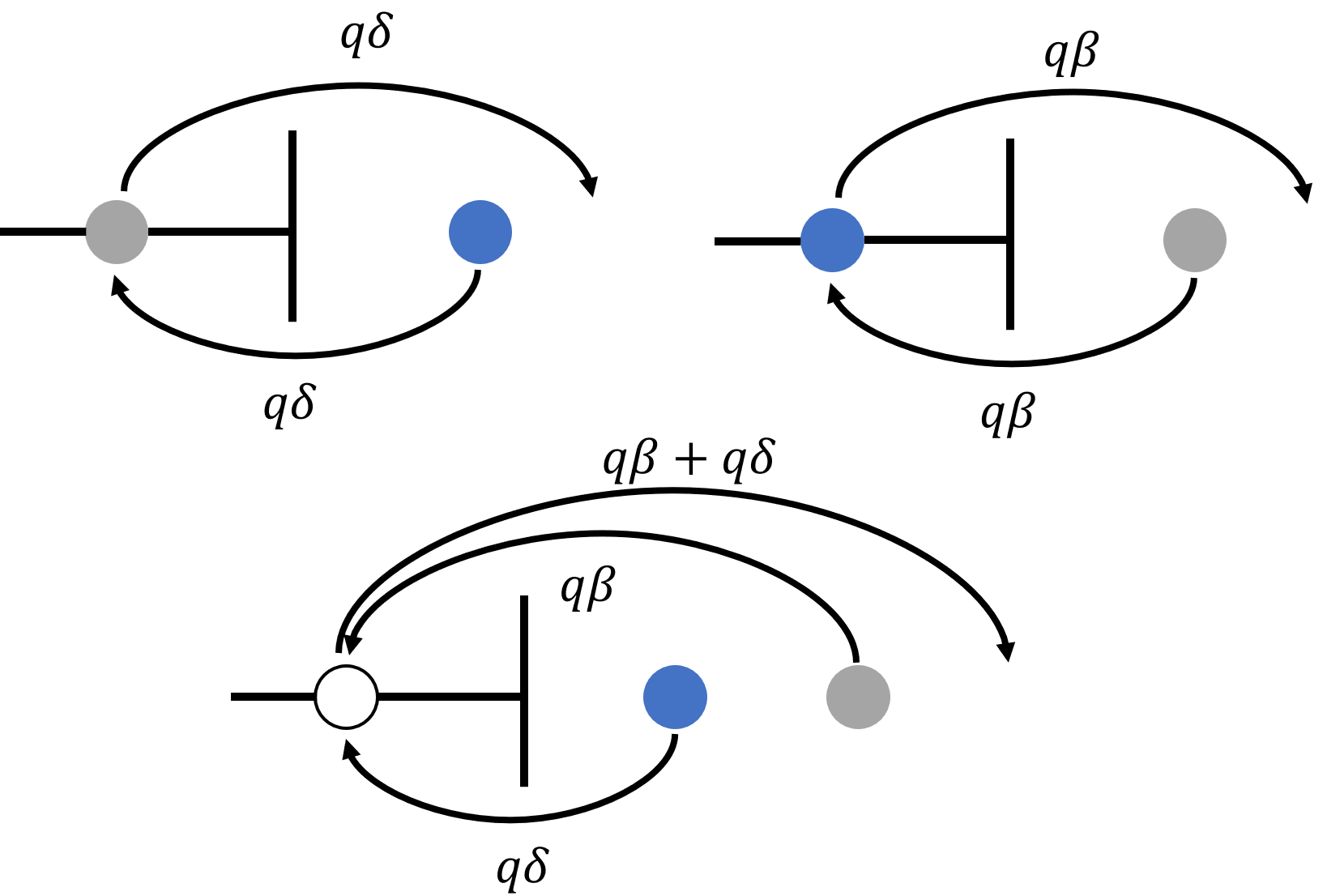}\\
  \caption{The transition rates at right boundary.}\label{figure-right-rate}
\end{figure}

We consider a one-dimensional lattice with $N$ sites and label particle configurations by strings $\nu=(\nu_1,\,\nu_2,\,\ldots,\,\nu_N)$ where $\nu_i\in\{-1,\,0,\,1\}$ and each label represents a particular species of particles. In this section, we focus on an open 2-ASEP with the following transition rates
\be
\begin{aligned}
&(\ldots,\nu_i,\nu_{i+1},\ldots)\stackrel{q}{\longrightarrow}(\ldots,\nu_{i+1},\nu_i,\ldots),\qquad \nu_i>\nu_{i+1},\quad\qquad\qquad~~~\\
&(\ldots,\nu_i,\nu_{i+1},\ldots)\stackrel{1}{\longrightarrow}(\ldots,\nu_{i+1},\nu_i,\ldots),\qquad \nu_i<\nu_{i+1},
\end{aligned}\\
(\nu_1,\ldots)\stackrel{q\gamma}{\longrightarrow}(-1,\ldots),\quad (-\nu_1,\ldots)\stackrel{q\alpha}{\longrightarrow}(1,\ldots),\quad\nu_1\in\{0,1\},\qquad~\label{left-rate}\\
(\ldots,\nu_N)\stackrel{q\beta}{\longrightarrow}(\ldots,-1),\quad (\ldots,-\nu_N)\stackrel{q\delta}{\longrightarrow}(\ldots,1),\quad\nu_N\in\{0,1\},\quad\label{right-rate}
\ee
where $q,\,\a,\,\b,\,\g$ and $\d$ are model parameters.
This 2-ASEP with open boundary conditions is depicted in Fig. \ref{figure-bulk-rate}, Fig. \ref{figure-left-rate} and Fig. \ref{figure-right-rate}

The time evolution equation of a state $|\Phi(t)\rangle$ for this 2-ASEP is given by
\be
\frac{{\rm d}}{{\rm d}t}|\Phi(t)\rangle=L|\Phi(t)\rangle,
\ee
where $L$ is the Markov matrix. In the tensor space
$\mathbf{V}_1\otimes\mathbf{V}_2\cdots\otimes \mathbf{V}_N$, we can write the Markov matrix in the following form
\be
L=L_1+\sum_{i=1}^{N-1}L_{i,i+1}+L_N,\label{markov-matrix}
\ee
where
\be\label{matrixform-L}
\begin{aligned}
&\quad~ L_1=\left(
       \begin{array}{ccc}
         -q\a & q\g & q\g \\
         0 & -q\a\!-\!q\g & 0 \\
         q\a & q\a & -q\g \\
       \end{array}
     \right),\\[4pt]
&\quad L_N=\left(
       \begin{array}{ccc}
         -q\d & q\b & q\b \\
         0 & -q\d\!-\!q\b & 0 \\
         q\d & q\d & -q\b \\
       \end{array}
     \right),\\[4pt]
&L_{i,i+1}=
\left({\scriptstyle{
              \begin{array}{ccc|ccc|ccc}
                0 & 0 & 0 & 0 & 0 & 0& 0 & 0 & 0 \\
                0 & -1 & 0& q & 0 & 0 &0 & 0 & 0 \\
                0 & 0 & -1& 0 & 0 & 0 & q & 0 & 0 \\
                \hline
                0 & 1 & 0 & -q & 0 & 0 & 0 & 0 & 0 \\
                0 & 0 & 0 & 0 & 0 & 0 &0 & 0 & 0 \\
                0 & 0 & 0 & 0 & 0 & -1 & 0 & q & 0 \\
                \hline
                0 & 0 & 1 & 0 & 0 & 0 & -q & 0 & 0 \\
                0 & 0 & 0 & 0 & 0 & 1 & 0 & -q & 0 \\
                0 & 0 & 0 & 0 & 0 & 0 & 0 & 0 & 0 \\
              \end{array}}}
            \right).
\end{aligned}
\ee
Here we adopt standard notation so that $L_1$ represents an operator acting in the full tensor product space but only non-trivially in $\mathbf{V_1}$ and as the identity on the other factors of the tensor product space; $L_N$ is an
operator acting non-trivially in the space $\mathbf{V_N}$, and as identity on the other factors;  $L_{i,i+1}$ is an operator acting non-trivially in the tensor space $\mathbf{V}_i\otimes \mathbf{V}_{i+1}$ and as identity on the other tensor spaces.

In the tensor space
$\mathbf{V}_1\otimes\mathbf{V}_2\cdots\otimes \mathbf{V}_N$, the sum of the elements in each column of Markov matrix is zero, meanwhile each row of the matrix has at least one non-zero element. So the left state $\langle\omega|\!=\!(1,1,\ldots,1)$ is an eigenvector of $L$ and $E_L\!=\!0$ is a non-degenerate eigenvalue of $L$. In addition, it is easy to check that $L^{\varepsilon,\ldots,\varepsilon}_{i_1,\ldots,i_N}
\!=\!-q(\a\!+\!\b\!+\!\g\!+\!\d)\prod_{k=1}^N\delta_{\varepsilon,i_k}$ with $\varepsilon=\frac{3^N\!+\!1}{2}$ where the superscript represents lines and subscripts represent rows. This indicates that $L$ also has an obvious eigenvalue $E_L\!=\!-q(\a\!+\!\b\!+\!\g\!+\!\d)$.

\subsection{Integrability}\label{integrability}
\subsubsection{$R$-matrix}
The $R$-matrix corresponding to the two-species ASEP is based on the universal $R$-matrix of the quantum group $U_{q^{1/2}}(A_2^{(1)})$, and reads
\be
{R}(x)=\left(
               \begin{array}{ccc|ccc|ccc}
                 a & 0 & 0 & 0 & 0 & 0 & 0 & 0 & 0 \\
                 0 & b^- & 0 & c^+ & 0 & 0 & 0 & 0 & 0 \\
                 0 & 0 & b^- & 0 & 0 & 0 & c^+ & 0 & 0 \\
                 \hline
                 0 & c^- & 0 & b^+ & 0 & 0 &0 & 0 & 0 \\
                 0 & 0 & 0 & 0 & a & 0 & 0 & 0 & 0 \\
                 0 & 0 & 0 & 0 & 0 & b^- & 0 & c^+ & 0 \\
                 \hline
                 0 & 0 & c^- & 0 & 0 & 0 & b^+ & 0 & 0\\
                 0 & 0 & 0 & 0 & 0 & c^- & 0 & b^+ & 0 \\
                 0 & 0 & 0& 0 & 0 & 0 & 0 & 0 & a \\
               \end{array}
             \right).\label{R-two-species}
\ee
Here we suppress the spectral parameter $x$ and $a(x)$, $b^{\pm}(x)$ and $c^{\pm}(x)$ are some functions of $x$ defined by
\begin{equation}
\begin{split}
&a(x)\!=\!q-x,\qquad b^+(x)\!=\!q-qx,\quad b^-(x)\!=\!1-x,\\
&c^-(x)\!=\!q-1,\qquad c^+(x)\!=\!qx-x.
\end{split}
\end{equation}
The $R$-matrix (\ref{R-two-species}) satisfies the Yang-Baxter equation (YBE)
\cite{yang1967,baxter1982,korepin1997}
\be
{R}_{12}(x_1/x_2){R}_{13}(x_1/x_3){R}_{23}(x_2/x_3)
={R}_{23}(x_2/x_3){R}_{13}(x_1/x_3){R}_{12}(x_1/x_2),\label{YBE}
\ee
and possesses the following properties
\be
&&{\mbox{Initial condition:}}\qquad \quad {R}_{12}(1)=(q-1){P}_{12},\label{initial-condition}\\
&&{\mbox{Unitary relation:}}\qquad\quad  {R}_{12}(x){R}_{21}(1/x)=\rho_1(x)\times \mathbb{I},\label{unitary-relation}\\
&&{\mbox{Crossing unitary relation:}}\quad\qquad  U^{-1}_{2}{R}^{t_1}_{12}(x)U_2{R}^{t_1}_{21}(q^3/x)\no\\
&&\hspace{5.5cm}=U^{-1}_{1}{R}^{t_1}_{12}(x)U_1{R}^{t_1}_{21}(q^3/x)=
\rho_2(x)\times \mathbb{I},\qquad\label{crossing-unitary-condition}
\ee
where $\mathbb{I}$ is the identity matrix, $P$ is the permutation matrix, ${R}_{21}(x)\!=\!{P}_{12}{R}_{12}(x){P}_{12}$, $\rho_1(x)\!=\!a(x)a(1/x)$, $\rho_2(x)\!=\!b^-(x)b^+(q^3/x)$ and $U\!=\!{\rm diag}\{1/q,\,1,\,q\}$.

\subsubsection{$K$-matrices}

For open systems, the integrability is guaranteed by the YBE and reflection equation, where the latter accounts for integrable boundaries \cite{sklyanin1988,cherednik1984}.
The two boundary reflection matrices of the model described above are the third class of Markovian $K$-matrices in \cite{Crampe}
\be
&&K^-(x)=
\left(\begin{array}{ccc}
{q(\g\hspace{-0.08cm}-\hspace{-0.08cm}\a)x^2\hspace{-0.08cm}+\hspace{-0.08cm}\eta_1x}
& {q\g (x^2\hspace{-0.08cm}-\hspace{-0.08cm}1)} & {q\g (x^2\hspace{-0.08cm}-\hspace{-0.08cm}1)} \\[4pt]
0 & -q\a x^2\hspace{-0.08cm}+\hspace{-0.08cm}\eta_1x\hspace{-0.08cm}+\hspace{-0.08cm}q\gamma & 0 \\[4pt]
{q\a (x^2\hspace{-0.08cm}-\hspace{-0.08cm}1)} & {q\a (x^2\hspace{-0.08cm}-\hspace{-0.08cm}1)} & \eta_1x\hspace{-0.08cm}+\hspace{-0.08cm}q(\g\hspace{-0.08cm}-\hspace{-0.08cm}\a)\hspace{-0.08cm}\\
\end{array}
\right),\label{K--1}\\[4pt]
&&K^{+}(x)=
\left(\begin{array}{ccc}
\eta_2 x \hspace{-0.08cm}+\hspace{-0.08cm}q^2(\d\hspace{-0.08cm}-\hspace{-0.08cm}q\b)\hspace{-0.08cm}& \b(x^2\hspace{-0.08cm}-\hspace{-0.08cm}q^3) & \b(x^2\hspace{-0.08cm}-\hspace{-0.08cm}q^3) \\[4pt]
0 & -q\b x^2\hspace{-0.08cm}+\hspace{-0.08cm}q\eta_2x\hspace{-0.08cm}+\hspace{-0.08cm}q^3\d\hspace{-0.08cm}& 0 \\[4pt]
 q\d (x^2\hspace{-0.08cm}-\hspace{-0.08cm}q^3) & q\d(x^2\hspace{-0.08cm}-\hspace{-0.08cm}q^3) & q(\d\hspace{-0.08cm}-\hspace{-0.08cm}q\b)x^2\hspace{-0.08cm}+\hspace{-0.08cm}q^2\eta_2x \\\end{array}\right).\label{K+-1}
\ee
where
\be
\eta_1=1-q+q\a-q\g,\quad \eta_2=1-q+q\b-q\d.\label{def-eta}
\ee
The matrices $ K^-(x)$ and $K^+(x)$ satisfy the following reflection equation (RE) and dual RE respectively
\be
&&{R}_{12}(x_1/x_2){K}_1^-(x_1){R}_{21}(x_1x_2){K}_2^-(x_2)
={K}_2^-(x_2){R}_{12}(x_1x_2){K}_1^-(x_1)R_{21}(x_1/x_2),\qquad\label{RE}\\[2pt]
&&{R}_{12}(x_1/x_2){K}_1^+(x_2)\widetilde{{R}}_{12}(x_1x_2){K}_2^+(x_1)
={K}_2^+(x_1)\widetilde{{R}}_{21}(x_1x_2){K}_1^+(x_2){R}_{21}(x_1/x_2),\qquad\label{DRE}
\ee
where $\widetilde{{R}}_{12}(x)=\rho_2(x)((R^{t_1}_{12}(x))^{-1})^{t_1}$.
The $K$-matrices $K^{\pm}(x)$ possess the following properties which will be useful later on,
\begin{equation}
\begin{aligned}\label{property-K}
&K^-(\pm1)=\left(q\g\pm\eta_1-q\a\right)\times \mathbb{I},\\
&K^+(\pm q^{\frac32})=\left(q^3\d\pm q^{\frac52}\eta_2-q^4\b\right)\times U,\\
&K^-(x)K^-(1/x)=h_1(x)\times \mathbb{I},\\
&K^+(x)U^{-1}K^-(q^3/x)=h_2(x)\times U,
\end{aligned}
\end{equation}
where
\begin{equation}
\begin{aligned}\label{function-h}
&h_1(x)=\left(q\a x^2\!-\!\eta_1 x\!-\!q\gamma\right)\left(q\a /x^2\!-\!\eta_1/x\!-\!q\gamma\right),\\
&h_2(x)=\left(q\b x^2\!-\!q\eta_2x\!-\!q^3\d\right)\left(q^7\b/ x^2\!-\!q^4\eta_2/x\!-\!q^3\d\right).
\end{aligned}
\end{equation}

Due to the fact that $[R_{12}(x),\, O_1O_2]\!=\!0$, where $O$ is a $3\times3$ diagonal constant matrix, the conjugated $K$-matrix $OK^-(x)O^{-1}$ also satisfies the reflection equation. The new system corresponding to the $K$-matrix $OK^-(x)O^{-1}$ is not a stochastic process and will contain some current generating function variables at the boundaries if $O\neq\mathbb{I}$.

\subsubsection{Transfer matrix}

The m-ASEP Markov generator $L$ is the logarithmic derivative of the transfer matrix.
In order to construct the transfer matrix,
we first define the following one-row monodromy matrices
\be
&&T_0(x)=R_{0N}(x/\theta_N)\cdots R_{01}(x/\theta_1),\label{T}\\
&&\widehat T_0(x)=R_{10}(x\theta_1)\cdots R_{N0}(x\theta_N),\label{hat-T}
\ee
where $\{\theta_1,\ldots,\theta_N\}$ are site-dependent inhomogeneous parameters. The transfer matrix then is given by
\be
\tau(x)={\rm tr}_0\left\{{K}_0^+(x){T}_0(x)K_0^-(x)\widehat T_0(x)\right\}.\label{transfer matrix}
\ee

Obviously, the transfer matrix is a sum of several operators which act on the tensor product space
$\mathbf{V}_1\otimes\mathbf{V}_2\cdots\otimes \mathbf{V}_N$.
The YBE (\ref{YBE}), RE (\ref{RE}) and dual RE (\ref{DRE}) lead to the fact that the transfer
matrices with different spectral parameters commute with each other, i.e., $[\tau(x),\tau(y)]=0$.
The Markov matrix $L$ is obtained as the logarithmic derivate of the transfer matrix $\tau(x)$ in the following way
\be
L=\frac{(1\hspace{-0.08cm}-\hspace{-0.08cm}q)}{2}\left.\frac{\partial\ln\tau(x)}{\partial x}\right|_{x=1,\{\theta_j=1\}}\!-\!N\!-\!\frac{q}{2}(\a\!+\!\b\!+\!\g\!+\!\d)\!-\!
\frac{1\!-\!2q\!+\!q^4}{1\!-\!q^3}.\label{L&tau}
\ee

\section{The fusion procedure}\label{nested bethe ansatz}\setcounter{equation}{0}

Our aim is to contruct a $T$-$Q$ relation \cite{baxter1982} for the open 2-ASEP using certain operator identities, as is done in the ODBA method. For the single-species ASEP with open boundaries, the corresponding transfer matrix processes a crossing symmetry which greatly decreases the number of needed operator identities to construct the $T$-$Q$ relation. In this case we can thus find sufficient operator identities just based on the transfer matrix.

The crossing symmetry of transfer matrix is broken for the higher rank open ASEP  and the previous method fails. Instead, we can construct a set of commuting fused transfer matrices \cite{karowski1979,kulish1990,kirillov1987} which will allow us to obtain a recursive set of operator product identities.

\subsection{Projectors}
In order to follow the approach suggested in the previous section and construct fused transfer matrices we introduce necessary projectors in this section. First let us define the vectors
\be
|i_1,i_2\ldots,i_n\rangle=|i_1\rangle\otimes |i_2\rangle\otimes\cdots\otimes|i_n\rangle,\quad i_k\!=\!1,2,3,\quad n\!=\!1,2,3,
\ee
where $|1\rangle\!=\!(1,\,0,\,0)^t$, $|2\rangle\!=\!(0,\,1,\,0)^t$ and $|3\rangle\!=\!(0,\,0,\,1)^t$.
We define furthermore the vectors
\be
|\phi_{i,j}\rangle=\frac{1}{\sqrt 2}(|i,j\rangle-|j,i\rangle),\qquad 1\!\leq\! i\!<\!j\!\leq\!3,
\ee
in the tensor space $\mathbf{V}\otimes \mathbf{V}$ and
\be
|\phi_{1,2,3}\rangle\!=\!\frac{1}{\sqrt6}\left(|1,2,3\rangle\!-\!|1,3,2\rangle\!+\!|3,1,2\rangle\!-\!|2,1,3\rangle
\!+\!|2,3,1\rangle\!-\!|3,2,1\rangle\right),
\ee
in the tensor space $\mathbf{V}\otimes \mathbf{V}\otimes \mathbf{V}$. Then, we can construct the
projection operators \cite{li2016}
\be
\begin{aligned}\label{projectors}
&P_{12}^-=\sum_{i<j}|\phi_{i,j}\rangle\langle\phi_{j,i}|,\\
&P_{123}^-=|\phi_{1,2,3}\rangle\langle\phi_{1,2,3}|.
\end{aligned}
\ee
We list a few properties of these projectors that will be useful,
\be
\begin{aligned}\label{Pr-P}
&P_{12}^-P_{12}^-=P_{12}^-,\quad P_{123}^-P_{123}^-=P_{123}^-,\\
&P_{12}^-=R_{12}(q)\,S_{12},\quad P_{123}^-=R_{12}(q)R_{13}(q^2)R_{23}(q)\,S_{123},
\end{aligned}
\ee
where
\be
\begin{aligned}
&S_{12}=(q\!-\!1)\!\times\!{\rm diag}\{0,\,1,\,1,\,1/q,\,0,\,1,\,1/q,\,1/q,\,0\},\\
&S_{123}=(q\!-\!1)^2(q^2\!-\!1)\!\times\!{\rm diag}\{0,\,0,\,0,\,0,\,0,\,1,\,0,\,1/q,\,0,\,0,\,0,\,1/q,\\
&\qquad\quad~~  0,\,0,\,0,\,1/q^2,\,0,\,0,\,0,\,1/q^2,\,0,\,1/q^3,\,0,\,0,\,0,\,0,\,0\}.
\end{aligned}
\ee
\subsection{Fused transfer matrix}

With the use of the projectors \eqref{Pr-P} we can construct the fused one-row monodromy matrices \cite{cao2014su(n)}
\be
&&T_{\langle12\ldots m\rangle}(x)=P_{mm-1\ldots 1}^-T_1(x)T_2(qx)\cdots T_m(q^{m-1}x)P_{mm-1\ldots1}^-,\\
&&\widehat T_{\langle12\ldots m\rangle}(x)=P_{12\ldots m}^-\widehat T_1(x)\widehat T_2(qx)\cdots \widehat T_m(q^{m-1}x)P_{12\ldots m}^-,
\ee
where the one-row monodromy matrices $T(x)$ and $\widehat T(x)$ are defined by (\ref{T}) and (\ref{hat-T}). For the higher rank model with periodic boundary conditions, the trace of $T_{\langle12\ldots m\rangle}(x)$ gives the fused transfer matrix. For the open boundary system we also need fusion for the $K$-matrices \cite{mezincescu1992,zhou1996}. We therefore introduce the following fused $K$-matrices
\be
&&K^-_{12\ldots m}(x)=K_1^-(x)R_{21}(qx^2)\cdots R_{m1}(q^{m-1}x^2)K_{\langle2\ldots m\rangle}^-(qx),\no\\
&&K^-_{\langle 12\ldots m\rangle }(x)=P^-_{mm-1\ldots1}K^-_{12\ldots m}(x)P_{12\ldots m}^-,\\
&&K^+_{12\ldots m}(x)=K^+_{\langle2\ldots m\rangle}(qx)\widetilde R_{m1}(q^{m-1}x^2)\cdots \widetilde R_{21}(qx^2)K_1^+(x),\no\\
&&K^+_{\langle 12\ldots m\rangle }(x)=P_{12\ldots m}^-K^+_{12\ldots m}(x)P_{mm-1\ldots1}^-.
\ee
where the boundary reflection matrices $K^{\pm}(x)$ are given by (\ref{K--1}) and (\ref{K+-1}). We are now in a position to construct the fused double row transfer matrix
\be
\tau_{m}(x)={\rm tr}_{12\ldots m}\left\{K^+_{\langle 1\ldots m\rangle}(x)T_{\langle 1\ldots m\rangle}(x)
K^-_{\langle 1\ldots m\rangle}(x)\widehat T_{\langle 1\ldots m\rangle}(x)\right\},\quad m=1,2,3.
\ee
Using the YBE, RE  and dual RE repeatedly we can prove that
\be
\left[\tau_{j}(x),\, \tau_{k}(y)\right]=0,\quad j,k=1,2,3.\label{commutation relation}
\ee
Here we use the notation $\tau_1(x)=\tau(x)$. The commutativity of the nested transfer matrices $\{\tau(x),\, \tau_2(x),\,\tau_3(x)\}$ implies that they share the same eigenvectors. If we therefore find certain operator identities between the nested transfer matrices, these identities immediately lift to functional relations.

\section{Operator identities}\label{operator identities}\setcounter{equation}{0}
The Yang-Baxter and reflection equations for integrable systems imply certain functional relations for the transfer matrix and fused transfer matrices. A direct consequence of these relations is that certain operator identities can be derived when the spectral parameter $x$ in the transfer matrix takes some special values.

Following the method in \cite{cao2014su(n)}, we arrive at the following recursive operator product identities for the fused transfer matrices $\{\tau(x),\,\tau_2(x),\,\tau_3(x)\}$
\be
&&\tau(\theta_j^{\pm1})\,\tau_m(q\theta_j^{\pm1})={\tau_{m+1}(\theta_j^{\pm 1})}{\prod_{k=1}^m\rho_2^{-1}(q^k\theta_j^{\pm2})},\quad j=1,\ldots,N,\quad m=1,2,\label{production-1}
\ee
The fused transfer matrix $\tau_3(x)$ is proportional to the identity matrix $\tau_3(x)\!=\!\Delta_q(x)\times\mathbb{I}$ where
\be
\Delta_q(x)&\hspace{-0.2cm}=&\hspace{-0.2cm}q^{-2N+6}\frac{(1\!-\!q^4x^2)(q^3\!-\!x^2)}{(1\!-\!qx^2)(1-x^2)}z(x)
z(q^2x)z(q^3x)\rho_2(qx^2)\rho_2(q^2x^2)\rho_2(q^3x^2)\no\\
&&\hspace{-0.2cm}\times (q^2\a x^2\!-\!\eta_1x\!-\g)(q^2\b x^2\!-\!\eta_2x\!-\!\d)
(q\a x^2\!-\!\eta_1x\!-\!q\g )(q\b x^2\!-\!\eta_2x\!-\!q\d )\no\\
&&\hspace{-0.2cm}\times (q\g x^2\!+\!\eta_1x\!-\!q\a)(q\d x^2\!+\!\eta_2x\!-\!q\b).\label{Delta}
\ee
Thus the operator production identities (\ref{production-1}) form a closed system. The identies (\ref{unitary-relation}) and (\ref{Pr-P}) imply another set of relations
\be
\tau_2(\theta_j^{\pm 1}/q)=0,\qquad\quad j=1,\ldots,N.\label{production-2}
\ee

Using the properties of $R$-matrix (\ref{initial-condition})--(\ref{crossing-unitary-condition}) and $K^{\pm}$-matrices (\ref{property-K}),
the values of fused transfer matrices $\tau(x)$ and $\tau_2(x)$ at some special points can be calculated directly. For example, using the unitary relation of $R$-matrix (\ref{unitary-relation}) and the initial condition of $K^-(x)$, we can easily obtain the following operator identity
\be
\begin{aligned}
\tau(1)&={\rm tr}_0\left\{K_0^+(1)T_0(1)K_0^-(1)\widehat{T}_0(1)\right\}\\
&=\frac{q^3\!-\!1}{q\!-\!1}(q\gamma\!+\!\eta_1\!-\!q\a)(q\d\!+\!\eta_2\!-\!q\b)\prod_{j=1}^N\rho_1(\theta_j)
\times\mathbb{I}.
\end{aligned}
\ee
Several other operator identities are given in detail in Appendix B resulting from considering these special points,
\be\label{SP-for-tau}
\left\{
     \begin{array}{ll}
       x\!=\!\pm1,~\pm q^{\frac32}\quad & \hbox{for $\tau(x)$} \\
       x\!=\!\pm1,~\pm q^{\frac12},~\pm q,~\pm q^{-\frac12},~ \pm q^{-1},~\pm q^{\frac32} & \hbox{for $\tau_2(x)$.}
     \end{array}
   \right.
\ee

When $x\!\rightarrow\!\pm\infty$ or $x\!=\!0$, the $R$-matrix and $K^{\pm}$-matrices simplify. In these limits we obtain the asymptotic behavior of the fused transfer matrices $\tau(x)$ and $\tau_2(x)$. Define
\be
t_1=\lim_{x\rightarrow\pm\infty}\frac{\tau(x)}{x^{2N+4}},\qquad
t_2=\lim_{x\rightarrow\pm\infty}\frac{\tau_2(x)}{x^{4N+10}}.
\ee
After some tedious calculations we find the explicit expression for $t_1$ and $t_2$,
\be
&&t_1=q^2\gamma\delta+q^{N+1}\a\b {W}+q^{N+2}\a\b {W}^{-1}+ G,\label{ASY-1}\\
&&t_2=-q^{3N+9}\a\b\left(\g\d W+q\g\d W^{-1}+q^N\a\b +\overline G\right),\label{ASY-2}
\ee
where $W=\otimes_{j=1}^N w_j$ and $w={\rm diag}\{1,\,q,\,1\}$. The matrices $G$ and $\overline G$ are additional terms that do not contribute to the diagonalisation of $t_1$ and $t_2$. With the help of the commutation relation (\ref{commutation relation}) we can readily prove that
$\tau(x)$, $\tau_2(x)$, $t_1$ and $t_2$ are mutually commutating with each other and thus they have common eigenstates. More details for matrices $t_1$ and $t_2$ are shown in Appendix \ref{A.1}.

\section{Nested off-diagonal Bethe ansatz}\label{T-Q relation}\setcounter{equation}{0}
We are now in a position to derive $T$-$Q$ relations and Bethe ansatz equations using the nested off-diagonal Bethe ansatz method.
\subsection{Functional relations}\label{functional-relations}
Suppose that $|\Psi\rangle$ is a common eigenvalue of $\tau(x),\tau_2(x)$ and $\tau_3(x)$, i.e.,
\be
\tau_m(x)|\Psi\rangle=\Lambda_m(x)|\Psi\rangle,\quad m=1,2,3,
\ee
where $\Lambda_m(x)$ is the corresponding eigenvalue of $\tau_m(x)$.
Here we use the notation $\L_1(x)=\L(x)$.
The function $\L_3(x)$ can be obtained directly as $\Lambda_3(x)=\Delta_q(x)$ where $\Delta_q(x)$ is defined in \eqref{Delta}.
The function $\L(x)$ is a degree $2N\!+\!4$ polynomial of $x$, and can be completely determined by $2N\!+\!5$ independent conditions. The function $x^2\L_2(x)$ is a degree $4N+12$ polynomial of $x$ (because the elements of $\widetilde{R}(x)$ are not all polynomials, an overall factor $x^2$ is added), and thus can be completely determined by $4N\!+\!13$ conditions.

The identities (\ref{production-1}), (\ref{production-2}) readily lead to the following relations
\be
&&\Lambda(\theta_j^{\pm1})\Lambda_m(q\theta_j^{\pm1})={\Lambda_{m+1}(\theta_j^{\pm 1})}{\prod_{k=1}^m\rho_2^{-1}(q^k\theta_j^{\pm2})},\quad j=1,\ldots,N,\label{production-3}\\[2pt]
&&\Lambda_2(\theta_j^{\pm 1}/q)=0,\quad\qquad j=1,\ldots,N,\label{production-4}
\ee
The values of $\L(x)$ and $\L_2(x)$ at the special points
\be
\begin{aligned}\label{SP-Lambda}
&\L(\pm1),\quad \L(\pm q^{\frac32}),\quad \L_2(\pm1),\quad \L_2(\pm q^{\frac12}),\\
&\L_2(\pm q),\quad \L_2(\pm q^{-\frac12}),\quad \L_2(\pm q^{-1}),\quad \L_2(\pm q^{\frac32}),
\end{aligned}
\ee
are the same as those of the fused transfer matrices $\tau(x)$ and $\tau_2(x)$ given in (\ref{SP}).
Furthermore, the diagonalization of  $t_1$ and $t_2$ gives the following asymptotic behavior
\be
&&\lim_{x\rightarrow\pm\infty}\frac{\Lambda(x)}{x^{2N+4}}= q^2\gamma\delta+q^{N+M+1}\a\b+q^{N-M+2}\a\b,\label{ASY-L-1}\\[4pt]
&&\lim_{x\rightarrow\pm\infty}\frac{\Lambda_2(x)}{x^{4N+10}}= -q^{3N+9}\a\b\left(q^M\g\d+q^{1-M}\g\d+q^N\a\b\right),\label{ASY-L-2}
\ee
where $M\!\in\!\{0,1,\ldots,N\}$.
The functional relations (\ref{production-3}) and (\ref{production-4}), the values at special points (\ref{SP-Lambda}) and the asymptotic behaviors (\ref{ASY-L-1}) and (\ref{ASY-L-2}) provide us with sufficient conditions to determine the corresponding eigenvalues $\L(x)$ and $\L_2(x)$ completely.

\subsection{Homogeneous $T$-$Q$ relation}\label{T-Q}

\subsubsection{Type one}
For convenience, introduce the notations
\be
&&y_1(x)=\frac{x^2-q^3}{x^2-q},\quad y_2(x)=\frac{x^2-1}{x^2-q^2},\label{function-y}\\[2pt]
&&z(x)=\prod_{j=1}^N(x/\theta_j-q)(x\theta_j-q),\label{function-z}\\[2pt]
&&f(x)=(q\a x^2-\eta_1x-q\g)(q\b x^2-\eta_2 x-q\d),\label{function-f}
\ee
where we recall the definition of $\eta_{1,2}$ given in \eqref{def-eta},
\be
\eta_1=1-q+q\a-q\g,\quad \eta_2=1-q+q\b-q\d.
\ee
We first define the functions $Q_{1}(x)$ and $Q_2(x)$ which are polynomials of $x$ parameterized as
\be\label{Q-1}
\begin{aligned}
&Q_1(x)=\prod_{j=1}^{N\hspace{-0.03cm}+\hspace{-0.03cm}M\hspace{-0.03cm}-\hspace{-0.03cm}1}
(x/\lambda_j-1)(x\lambda_j-q),\\
&Q_2(x)=\prod_{k=1}^{M}(x/\mu_k-1)(x\mu_k-q^2),
\end{aligned}
\ee
in terms of the Bethe roots $\{\lambda_j\}$ and $\{\mu_k\}$ which are yet to be determined.

The functional identities in Section \ref{functional-relations} allow us to construct the following $T$-$Q$ relation.
\be
\begin{aligned}\label{TQ-1-1}
\L(x)=&~q^{-N-M+1}y_1(x)f(x)z(x)\frac{{Q}_1(qx)}{{Q}_1(x)}+q^{M-N+4}y_2(x)f(x/q)z(qx)\frac{{ Q}_2(x/q)}{{Q}_2(x)}\qquad~~~\\
&~+q^{-2}x^4y_1(x)y_2(x)f(q/x)z(qx)\frac{{Q}_1(x/q) {Q}_2(qx)}{{Q}_1(x){Q}_2(x)},
\end{aligned}\\
\begin{aligned}\label{TQ-1-2}
\L_2(x)=&~y_1(x)y_2(qx)\rho_2(qx^2)z(q^2x)
\left\{q^{-N-M+3}x^4y_1(qx)f(x)f(1/x)z(x)\frac{{Q}_2(q^2x)}{{Q}_2(qx)}\right.\qquad\\
&~+\left.q^{-2N+5}f(x)f(x)z(x)\frac{{Q}_1(qx){Q}_2(x)}{{Q}_1(x){Q}_2(qx)}\right.\\
&~+\left.q^{M-N+2}x^4y_2(x)f(q/x)f(x)z(tx)\frac{{Q}_1(x/q)}{{Q}_1(x)}\right\},\qquad\quad
\end{aligned}
\ee
where $M\!\in\!\{0,1,\ldots,N\}$. The regular property of $\L(x)$ and $x^2\L_2(x)$ induces the following Bethe ansatz equations(BAEs) for the Bethe roots $\{\lambda_j\}$ and $\{\mu_k\}$,
\be
\begin{aligned}\label{BE-1}
&q^{N+M-3}\,\,y_2(\l_k)\,\frac{f(q/\l_k)}{f(\l_k)}\frac{z(q\l_k)}{z(\l_k)}\frac{Q_1(\l_k/q)}{ Q_1(q\l_k) }\frac{Q_2(q\l_k)}{Q_2(\l_k)}\!=\!-1,\quad k\!=\!1,\ldots,N\!+\!M\!-\!1,\quad\\
&q^{N-M-6}\,\mu_j^4\,y_1(\mu_j)\,\frac{f(q/\mu_j)}{f(\mu_j/q)}\frac{Q_1(\mu_j/q)}{Q_1(\mu_j)}\frac{Q_2(q\mu_j)}{ Q_2(\mu_j/q)}\!=\!-1,\quad j\!=\!1,\ldots,M,
\end{aligned}
\ee
where the Bethe roots should satisfy $\l_j\!\neq\!\l_k\!\neq\! q/\l_l$, $\mu_j\!\neq\!\mu_k\!\neq\! q^2/\mu_l$ and $\l_j\!\neq \!\mu_k$.

The $T$-$Q$ relation (\ref{TQ-1-1}) can give the complete eigenvalues of transfer matrix $\tau(x)$. Some numerical results are given in Section \ref{numerical check}. The eigenvalues of Markov matrix $L$ in terms of Bethe roots $\{\l_j\}$ are recovered by setting $\{\theta_j\!=\!1\}$, and in this case are given by
\be
E_L=-\sum_{k=1}^{N\hspace{-0.03cm}+\hspace{-0.03cm}M\hspace{-0.03cm}-\hspace{-0.03cm}1}\frac{(q-1)^2}{(1-1/\l_k)(q-\l_k)}-q(\a+\b+\g+\d).
\ee

\subsubsection{Type two}

The $T$-$Q$ relation is not unique. Different $T$-$Q$ relations mean different parameterization of the functions $\Lambda(x)$ and $\L_2(x)$.
We can construct another $T$-$Q$ relation (here and below we omit the $T$-$Q$ relation for $\L_2(x)$)
\be
\begin{aligned}\label{TQ-2}
\L(x)=&~x^4y_1(x)f(1/x)z(x)+q^{M-N+4}y_2(x)f(x/q)z(qx)\frac{{\widetilde Q_2}(x/q)}{ \widetilde Q_2(x)}\\
&~+q^{-M-N+3}y_1(x)y_2(x)f(x/q)z(qx)\frac{\widetilde Q_2(qx)}{\widetilde Q_2(x)},
\end{aligned}
\ee
where
\be
{\widetilde Q_2}(x)=\prod_{k=1}^{M}(x/\widetilde \mu_k-1)(x\widetilde \mu_k-q^2).
\ee
The Bethe roots $\{\widetilde\mu_j\}$ need to satisfy the selection rules $\widetilde \mu_j\!\neq\!\widetilde\mu_k\!\neq\! q^2/\widetilde\mu_l$ and the following BAEs for $\L(x)$ to be regular,
\be
q^{-2M-1}\,y_1(\widetilde\mu_j) \frac{\widetilde Q_2(q\widetilde\mu_j)}{{\widetilde Q_2}(\widetilde\mu_j/q)}=-1,\quad j=1,\ldots,M.\label{BE-2}
\ee

The expressions of $\Lambda(x)$ are equal for $M\!=\!0$ and $M\!=\!1$. When $M\!\geq\!2$, the BAEs (\ref{BE-2}) have no solutions. So the $T$-$Q$ relation (\ref{TQ-2}) only gives one eigenvalue of $\tau(x)$, namely the one when $M\!=\!0$. After some simple calculations, the corresponding eigenvalue of the Markov matrix is
$E_L=0$.

\subsubsection{Type three}

Yet another alternative $T$-$Q$ relation is
\be
\begin{aligned}\label{TQ-3}
\L(x)=&~q^{-\overline M}y_1(x)f(x)z(x)\frac{{\overline Q}_1(qx)}{{\overline Q}_1(x)}+q^{-4}x^4y_2(x) f(q^2/x)z(qx)\frac{{\overline Q}_2(x/q)}{{\overline Q}_2(x)}\\
&~+q^{\overline M-2N+3}y_1(x)y_2(x) f(x)z(qx)\frac{{\overline Q}_1(x/q){\overline Q}_2(qx)}{{\overline Q}_1(x){\overline Q}_2(x)},
\end{aligned}
\ee
where $\overline M\!\in\!\{0,1,2\ldots,N\!\!-\!\!1\}$.
The $Q$-functions are defined by
\be
\begin{aligned}\label{Q-2}
&{\overline Q}_1(x)=\prod_{k=1}^{\overline M}(x/\overline \lambda_k-1)(x\overline \lambda_k-q),\\
&{\overline Q}_2(x)=\prod_{k=1}^{N\!-\!2}(x/\overline \mu_k-1)(x\overline\mu_k-q^2).
\end{aligned}
\ee
The Bethe roots are the solution of the BAEs
\be
\begin{aligned}\label{BE-3}
&q^{2\overline M-2N+3}\,y_2(\overline\l_k)\,\frac{z(q\overline\l_k)}{z(\overline\l_k)}\frac{\overline Q_1(\overline\l_k/q)}{\overline Q_1(q\overline\l_k)}\frac{\overline Q_2(q\overline\l_k)}{\overline Q_2(\overline\l_k)}\!=\!-1,\qquad k\!=\!1,\ldots,\bar M,\\
&q^{\overline M-2N+7}\,{\overline\mu_j}^{-4}\,y_1(\overline\mu_j)\,\frac{f(\overline\mu_j)}{f(q^2/\overline \mu_j)}\frac{\overline Q_1(\overline\mu_j/q)}{\overline Q_1(\overline\mu_j)}
\frac{\overline Q_2(q\overline\mu_j)}{\overline Q_2(\overline\mu_j/q)}\!=\!-1,\qquad j\!=\!1,\ldots,N\!-\!2,
\end{aligned}
\ee
where $\overline\l_j\!\neq\!\overline\l_k\!\neq\! q/\overline\l_l$, $\overline\mu_j\!\neq\!\overline\mu_k\!\neq\! q^2/\overline\mu_l$ and
$\overline\l_j\!\neq\!\overline\mu_k$.
The eigenvalue of Markov matrix in terms of the Bethe roots is again recovered from setting $\{\theta_j\!=\!1\}$ and is given by
\be
E_L=-\sum_{k=1}^{\overline M}\frac{(q-1)^2}{\left(1-1/\overline\l_k\right)\left(q-\overline\l_k\right)}-q(\a+\b+\g+\d).
\ee

The $T$-$Q$ relation (\ref{TQ-3}) does not give the complete set of eigenvalues either of transfer matrix $\tau(x)$. However, compared with (\ref{TQ-1-1}), the $T$-$Q$ relation (\ref{TQ-3}) is more concise when we want to parameterize some particular $\L(x)$.
 For instance, when $\overline M\!=\!0$, the $T$-$Q$ relation (\ref{TQ-3}) directly gives $E_L\!=\!-q(\a+\b+\g+\d)$, while in \eqref{TQ-1-1} one always has to consider at least $N\!-\!1$ Bethe roots. Some numerical results for the $T$-$Q$ relation \eqref{TQ-3} are also given in Section \ref{numerical check}.

\subsection{Numerical result}\label{numerical check}
\setcounter{figure}{0}
\setcounter{table}{0}

Let $\{\theta_j\!=\!1\},\,q=4.0,\,\a=1.2,\,\b=2.4,\,\g=3.5$ and $\d=7.0$. We have performed the following numerical check for $N\!=\!2$ case.
The numerical solutions of nested BAEs (\ref{BE-1}) and (\ref{BE-3})
are shown in Table \ref{table-1} and Table \ref{table-2} respectively. In order to verify our $T$-$Q$ relations,
we also calculate the eigenvalues of Markov matrix $E_L$ in terms of the obtained Bethe roots.

\begin{table}[htp]
\begin{center}
\small
\begin{tabular}{|c|c|c|c|c|c|c|}
  \hline\rule{0pt}{12pt}
   $\lambda_1$ & $\lambda_2$ & $\l_3$& $\mu_1$& $\mu_2$& $E_L$  & $d$\\[2pt]
 \hline 1.8425$+$0.7780$i$ & 0.4229& & 3.6309$+$1.6782$i$ & &$-$61.4000  & 1\\
 \hline $-$0.1633$-$1.9933$i$ & & & & &$-$58.0897   & 2\\
\hline 0.1467$-$0.5508$i$& 0.3181 & 1.8063$-$6.7815$i$ & 1.0928 &0.3793& $-$56.4000  & 1\\
 \hline0.8561 & & & & & $-$39.3654  & 2\\
 \hline0.9249 & & & & & $-$20.3449  & 2\\
 \hline 0.8843& 0.9215 & &$-$0.1784 & & 0.0000  & 1\\
  \hline
\end{tabular}
\caption{The numerical solutions of  BAEs (\ref{BE-1}). The symbol $d$ indicate the number of degeneracy.}\label{table-1}
\end{center}

\end{table}

\begin{table}[htp]
\begin{center}
\small
\begin{tabular}{|c|c|c|}
  \hline\rule{0pt}{12pt}
   $\overline \lambda_1$ &  $E_L$  &$d$\\[2pt]
 \hline 1.6000$+$1.2000$i$&   $-$61.4000 & 1\\
 \hline &   $-$56.4000 & 1\\
  \hline
\end{tabular}
\end{center}
\caption{The numerical solutions of BAEs (\ref{BE-3}).}\label{table-2}
\end{table}
\newpage

\begin{figure}[htp]
  \centering
  \includegraphics[width=11cm]{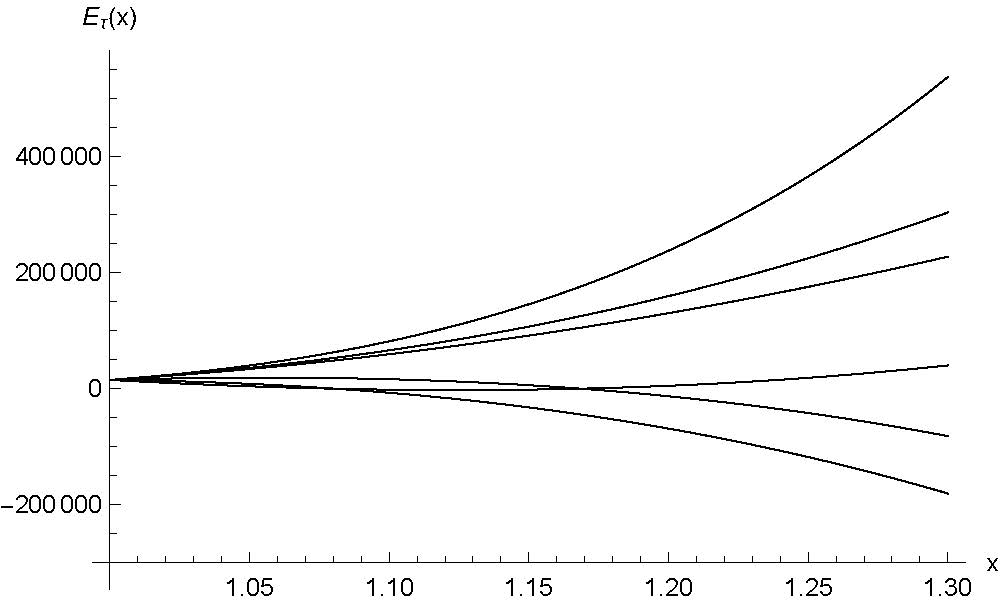}\\
  \caption{The eigenvalue curves of transfer matrix $\tau(x)$ obtained from the exact diagonalization of $\tau(x)$.}\label{figure-1}
\end{figure}

\begin{figure}[htp]
  \centering
  \includegraphics[width=13cm]{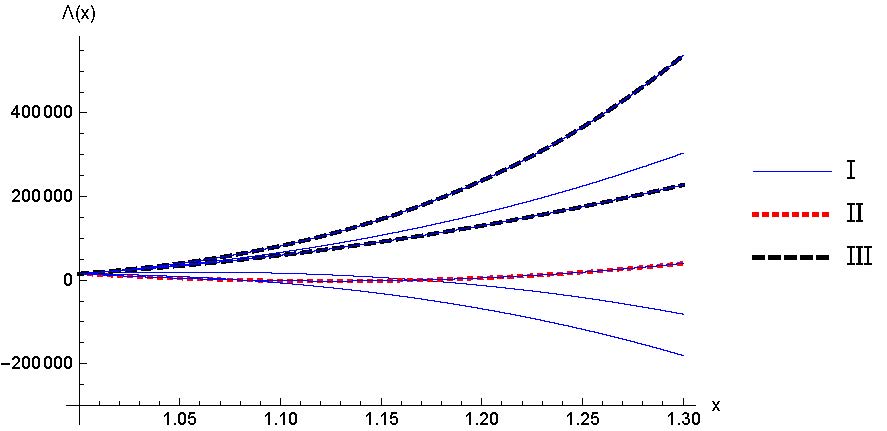}\\
  \caption{The eigenvalue curves of transfer matrix $\tau(x)$ given by the $T$-$Q$ relations where ``I'', ``II'', ``III'' represent $T$-$Q$ relations (\ref{TQ-1-1}), (\ref{TQ-2}) and (\ref{TQ-3}) respectively.}\label{figure-2}
\end{figure}

The eigenvalues of Markov matrix parameterized by Bethe roots are consistent
with the exact diagonalization of Markov matrix $L$ (\ref{markov-matrix}).
A more straightforward verification of our results is to compare our $T$-$Q$ relations and the exact diagonalization of transfer matrix $\tau(x)$ (\ref{transfer matrix}). In Fig. \ref{figure-1} and Fig. \ref{figure-2} we show that the curves calculated from $T$-$Q$ relation and the nested BAEs are exactly the same as those obtained from the exact diagonalization of the transfer matrix $\tau(x)$.

\section{Another integrable 2-ASEP}\label{another-K}\setcounter{equation}{0}

In this section we consider another, more frequently studied integrable 2-ASEP with open boundaries which is described by a new Markov matrix
\be
\overline L=\overline L_1+\sum_{i=1}^{N-1}L_{i,i+1}+\overline L_N,\label{new-markov-matrix}
\ee
where $L_{i,i+1}$ is given by (\ref{matrixform-L}) and
\be
\begin{aligned}\label{new-boundary-condition}
&\overline L_1=\left(
           \begin{array}{ccc}
            -q\a & 0 & q\g \\
             0 & 0 & 0 \\
            q\a & 0 & -q\g \\
           \end{array}
         \right),\\[4pt]
&\overline L_N=\left(
           \begin{array}{ccc}
             -q\d & 0 & q\b \\
             0 & 0 & 0 \\
             q\d & 0 & -q\b \\
           \end{array}
         \right).
\end{aligned}
\ee
The transition rates of this models can be directly observed from the matrix form of $\overline L$
\be
\begin{aligned}\label{new-bulk-rate}
&(\ldots,\nu_i,\nu_{i+1},\ldots)\stackrel{q}{\longrightarrow}(\ldots,\nu_{i+1},\nu_i,\ldots),\quad \nu_i>\nu_{i+1},\\
&(\ldots,\nu_i,\nu_{i+1},\ldots)\stackrel{1}{\longrightarrow}(\ldots,\nu_{i+1},\nu_i,\ldots),\quad \nu_i<\nu_{i+1},
\end{aligned}\\
(1,\ldots)\stackrel{q\gamma}{\longrightarrow}(-1,\ldots),\qquad (-1,\ldots)\stackrel{q\alpha}{\longrightarrow}(1,\ldots),~~~~~~\label{new-left-rate}\\
(\ldots,1)\stackrel{q\beta}{\longrightarrow}(\ldots,-1),\qquad (\ldots,-1)\stackrel{q\delta}{\longrightarrow}(\ldots,1).~~~~\label{new-right-rate}
\ee

\subsection{Integrability}
The boundary $K$-matrices corresponding to this model are given by \cite{Crampe}:
\be
&&\overline K^-(x)=\left(
                \begin{array}{ccc}
                  q(\g\hspace{-0.05cm}-\hspace{-0.05cm}\a)x^2\hspace{-0.05cm}+\hspace{-0.05cm}\eta_1x & 0 & q\g(x^2\hspace{-0.05cm}-\hspace{-0.05cm}1) \\
                  0 & q\g x^2\hspace{-0.05cm}+\hspace{-0.05cm}\eta_1 x\hspace{-0.05cm}-\hspace{-0.05cm}q\a & 0 \\
                  q\a(x^2\hspace{-0.05cm}-\hspace{-0.05cm}1) & 0 & \eta_1x\hspace{-0.05cm}+\hspace{-0.05cm}q(\g\hspace{-0.05cm}-\hspace{-0.05cm}\a) \\
                \end{array}
              \right)
,\label{NK-}\\[4pt]
&&\overline K^+(x)=\left(
                \begin{array}{ccc}
                  \eta_2x\hspace{-0.05cm}+\hspace{-0.05cm}q^2(q\d\hspace{-0.05cm}-\hspace{-0.05cm}\b) & 0 & \b(x^2\hspace{-0.05cm}-\hspace{-0.05cm}q^3) \\
                  0 & q\d x^2\hspace{-0.05cm}+\hspace{-0.05cm}q\eta_2x\hspace{-0.05cm}-\hspace{-0.05cm}q^3\b & 0 \\
                  q\d(x^2\hspace{-0.05cm}-\hspace{-0.05cm}q^3) & 0 & q(q\d\hspace{-0.05cm}-\hspace{-0.05cm}\b)x^2\hspace{-0.05cm}+\hspace{-0.05cm}q^2\eta_2x \\
                \end{array}
              \right).\label{NK+}
\ee
We can easily prove that $\overline K^-(x)$ (\ref{NK-}) is consistent with the Markovian $K$-matrix in \cite{cantini2016} if we switch the parameters as $\a=\frac{(q-1)\bar a\bar c}{(1+\bar a)(1+\bar c)q}$, $\g=\frac{(1-q)}{(1+\bar a)(1+\bar c)q}$.
The new $K$-matrices $\overline K^-(x)$ and $\overline K^+(x)$ satisfy the RE (\ref{RE}) and dual RE (\ref{DRE}) respectively. The commuting transfer matrix $\overline\tau(x)$ is constructed as
\be
\overline\tau(x)={\rm tr}_0\left\{\overline K_0^+(x)T_0(x)\overline K_0^-(x)\widehat T_0(x)\right\}.\label{New-t}
\ee
The Markov matrix $\overline L$ can be given by the transfer matrix $\overline\tau(x)$ as follow
\be
\overline L=\frac{(1\!-\!q)}{2}\left.\frac{\partial\ln\overline\tau(x)}{\partial x}\right|_{x=1,\{\theta_j=1\}}
\!-\!N\!-\!\frac{q}{2}(\a\!+\!\b\!+\!\g\!+\!\d)\!-\!
\frac{1\!-2q\!+q^4}{1\!-\!q^3}.\label{new-L&tau}
\ee

\subsection{$T$-$Q$ relation}
\subsubsection{Type one}

Assume that $\overline\Lambda(x)$ is an eigenvalue of $\overline\tau(x)$. Using the same procedure as above we can construct the following homogeneous $T$-$Q$ relation for this model
\be
\begin{aligned}\label{new-TQ-1}
\overline\Lambda(x)=&~q^{M-N} x^4y_1(x)f(1/x)z(x)\frac{
\mathcal{Q}_1(qx)}{\mathcal{Q}_1(x)}+x^4y_2(x)f(q/x)z(qx)
\frac{\mathcal Q_2(x/q)}{\mathcal Q_2(x)}\\
&~+q^{-N-M+3}y_1(x)y_2(x)f(x/q)z(qx)
\frac{\mathcal Q_1(x/q)\mathcal Q_2(qx)}{\mathcal Q_1(x)\mathcal Q_2(x)},
\end{aligned}
\ee
where $M\!\in\!\{0,1,\ldots,N\}$ and the functions $y_1(x)$, $y_2$ and $f(x)$ are defined by
(\ref{function-y})-(\ref{function-f}).
The functions $\mathcal{Q}_1(x)$ and $\mathcal{Q}_2(x)$ are
\be
\begin{aligned}\label{new-Q-1}
&\mathcal{Q}_1(x)=\prod_{j=1}^{N\hspace{-0.05cm}-\hspace{-0.05cm}M}(x/{\lambda}_j-1)(x\lambda_j-q),\label{def-new-Q-1}\\
&\mathcal{Q}_2(x)=\prod_{k=1}^N(x/{\mu}_k-1)(x\mu_k-q^2).
\end{aligned}
\ee
The Bethe roots in (\ref{new-Q-1}) satisfy the following BAEs
\be
\begin{aligned}\label{new-BE-1}
&q^{-2M+3}\,\l_j^{-4}\,y_2(\lambda_j)\,\frac{f(\l_j/q)}{f(1/\l_j)}\frac{z(q\lambda_j)}{z(\lambda_j)}
\frac{\mathcal Q_1(\lambda_j/q)}{\mathcal Q_1(q\lambda_j)}\frac{\mathcal Q_2(q\lambda_j)}{\mathcal Q_2(\lambda_j)}\!=\!-1,\quad j\!=\!1,\ldots,N\!\!-\!\!M,\\
&q^{-N-M+3}\,\mu_k^{-4}\,y_1(\mu_k)\,\frac{f(\mu_k/q)}{f(q/\mu_k)}\frac{\mathcal Q_1(\mu_k/q)}{\mathcal Q_1(\mu_k)}\frac{\mathcal Q_2(q\mu_k)}{\mathcal Q_2(\mu_k/q)}\!=\!-1,\quad k\!=\!1,\ldots,N.
\end{aligned}
\ee
The selection rules for
these Bethe roots are $\l_j\!\neq\!\l_k\!\neq\!q/\l_l$, $\mu_j\!\neq\!\mu_k\!\neq\!q^2/\mu_l$ and $\l_j\!\neq\!\mu_k$ except for the special case when $\{\l_j\!=\!0\}$.
Let $\{\theta_j\!=\!1\}$, then
the eigenvalue of Markov matrix $\overline L$ in terms of Bethe roots is given by
\be
E_{\overline L}=-\sum_{j=1}^{N\!-\!M}\frac{(q-1)^2}{(q-\l_j)(1-1/\l_j)}
\ee

The total number of particles labeled by ``0'' is given by the integer $M$ which is conserved, and so the transfer matrix $\overline\tau(x)$ has an unbroken $U(1)$ symmetry.
The ``completely empty'' state, i.e. only particles of type ``0'', is an eigenstate of the transfer matrix $\overline\tau(x)$.
Therefore, we can carry out the first step of the nested algebraic Bethe ansatz
with this state as a reference state.
 The $T$-$Q$ relation (\ref{new-TQ-1}) can also be constructed using the nested algebraic Bethe ansatz and ODBA step by step. All the eigenvalues of transfer matrix $\overline\tau(x)$ can be parameterized by the $T$-$Q$ relation (\ref{new-TQ-1}). The Bethe eigenstates can then be constructed via the nested algebraic Bethe ansatz when we know the distribution of Bethe roots $\{\l_j\}$.

\subsubsection{Type two}
We can also construct another $T$-$Q$ relation for this model,
\be
\begin{aligned}\label{new-TQ-2}
\overline\L(x)=&~x^4y_1(x)f(1/x)z(x)+q^{M-N}x^4y_2(x)f(q/x)z(qx)
\frac{\mathcal{ \overline Q}_2(x/q)}{\mathcal {\overline Q}_2(x)}\\
&~+q^{-M-N+3}y_1(x)y_2(x)f(x/q)z(qx)
\frac{\mathcal{\overline Q}_2(qx)}{\mathcal{\overline Q}_2(x)},
\end{aligned}
\ee
where the function $\mathcal{\overline Q}_2(x)$ is parameterized by the Bethe roots as follows,
\be
&&\mathcal{\overline Q}_2(x)=\prod_{k=1}^M(x/{\overline \mu}_k-1)(x\overline \mu_k-q^2).
\ee
The Bethe roots $\{\overline\mu_j\}$ in (\ref{new-TQ-2}) are the solutions of the following BAEs
\be
q^{-2M+3}\,{\overline\mu}_j^{-4}\,y_1(\overline\mu_j)\,\frac{f(\overline\mu_j/q)}{f(q/\overline\mu_j)}
\frac{\mathcal{\overline Q}_2(q\overline\mu_j)}
{\mathcal{\overline Q}_2(\overline\mu_j/q)}\!=\!-1,\quad j\!=\!1,\ldots,M.\label{new-BE-2}
\ee
The selection rules for $\{\overline\mu_j\}$ are $\overline\mu_j\!\neq\!\overline\mu_k\!\neq\! q^2/\overline\mu_l$.
Using the identity (\ref{new-L&tau}), we can easily prove the $T$-$Q$ relation (\ref{new-TQ-2}) corresponds to $E_{\overline L}=0$. The further numerical results for small scale systems indicate that the $T$-$Q$ relation (\ref{new-TQ-2}) can give us all the message of the degenerate eigenvalue $E_{\overline L}=0$.

\subsubsection{Type three}

Yet another alternative $T$-$Q$ relation is
\be
\begin{aligned}\label{new-TQ-3}
&\overline\L(x)=q^{-N-M+1}y_1(x)f(x)z(x)\frac{
\mathcal{\widetilde Q}_1(qx)}{\mathcal{\widetilde Q}_1(x)}+q^{ M-N}x^{4}y_2(x)f(q/x)z(qx)
\frac{\mathcal{ \widetilde Q}_2(x/q)}{\mathcal {\widetilde Q}_2(x)}\\
&\hspace{1.5cm}+q^{-2}x^{ 4}y_1(x)y_2(x)f(q/x)z(qx)
\frac{\mathcal{\widetilde Q}_1(x/q)\mathcal{\widetilde Q}_2(qx)}{\mathcal{\widetilde Q}_1(x)\mathcal{\widetilde Q}_2(x)},
\end{aligned}
\ee
where the functions $\mathcal{\widetilde Q}_1(x)$ and $\mathcal{\widetilde Q}_2(x)$ are defined by
\be
\begin{aligned}\label{new-Q-3}
&\mathcal{\widetilde Q}_1(x)=\prod_{j=1}^{N\hspace{-0.04cm}+\hspace{-0.04cm}M\hspace{-0.04cm}-\hspace{-0.04cm}1}
(x/{\widetilde\lambda}_j-1)(x\widetilde\lambda_j-q),\\
&\mathcal{\widetilde Q}_2(x)=\prod_{k=1}^M(x/{\widetilde\mu}_k-1)(x\widetilde\mu_k-q^2).
\end{aligned}
\ee
The requirement that $\overline\L(x)$ should not have any poles leads to the following BAEs
\be
\begin{aligned}\label{new-BE-3}
&q^{N+M-3}\,{\widetilde\l}_j^4\,y_2(\widetilde\l_j)\,
\frac{f(q/\widetilde\l_j)}{f(\widetilde\l_j)}\frac{z(q\widetilde\l_j)}{z(\widetilde\l_j)}
\frac{\mathcal{\widetilde Q}_1(\widetilde\l_j/q)}{\mathcal{\widetilde Q}_1(q\widetilde\l_j)}
\frac{\mathcal{\widetilde Q}_2(q\widetilde\l_j)}{\mathcal{\widetilde Q}_2(\widetilde\l_j)}\!=\!-1,\quad j\!=\!1,\ldots,N\!+\!M\!-\!1,\\
&q^{N-M-2}\,y_1(\widetilde\mu_k)\,\frac{\mathcal{\widetilde Q}_1(\widetilde\mu_k/q)}{\mathcal{\widetilde Q}_1(\widetilde\mu_k)}\frac{\mathcal{\widetilde Q}_2(q\widetilde\mu_k)}{\mathcal{\widetilde Q}_2(\widetilde\mu_k/q)}\!=\!-1,\qquad\quad k\!=\!1,\ldots,M.
\end{aligned}
\ee
where Bethe roots should satisfy the selection rules $\widetilde\l_j\!\neq\!\widetilde\l_k\!\neq\! q/\widetilde\l_l$, $\widetilde\mu_j\!\neq\!\widetilde\mu_k\!\neq\! q^2/\widetilde\mu_l$ and $\widetilde\l_j\!\neq\!\widetilde\mu_k$.
Let $\{\theta_j\!=\!1\}$, the corresponding eigenvalue of Markov matrix $\overline L$ is
\be
E_{\overline L}=-\sum_{j=1}^{N\!+\!M\!-\!1}\frac{(q-1)^2}{(q-\widetilde\l_j)(1-1/\widetilde\l_j)}\!-\!q(\a\!+\!\b\!+\!\g\!+\!\d).
\ee

The numerical results in Section \ref{new-numerical results} show that the $T$-$Q$ relation (\ref{new-TQ-3}) can not parameterize all the eigenvalues of transfer matrix. However, less Bethe roots are used in (\ref{new-TQ-3}) to parameterize the function $\overline\L(x)$ in some cases compared with the $T$-$Q$ relation (\ref{new-TQ-1}), and so the eigenvalues that are included are in a more convenient form.

\subsection{Numerical results}\label{new-numerical results}
\setcounter{table}{0}
\setcounter{figure}{0}

Let $\{\theta_j\!=\!1\}$, $q\!=\!1.8$, $\a\!=\!0.22$, $\b\!=\!0.41$, $\g\!=\!0.76$ and $\d\!=\!0.95$, then we do the numerical check for the $N\!=\!2$ case. The numerical solutions of the BAEs
 (\ref{new-BE-1}), (\ref{new-BE-2}) and (\ref{new-BE-3})
for the $N\!=\!2$ case are shown in Table \ref{new-table-1}, Table \ref{new-table-2} and Table \ref{new-table-3} respectively.
\begin{table}[htp]
\begin{center}
\small
\begin{tabular}{|c|c|c|c|c|}
\hline\rule{0pt}{13pt}
   $\lambda_1$ & $\lambda_2 $ & $\mu_1$ &$\mu_2$ & $E_{\overline L}$\\[4pt]
   \hline 1.3364$-$0.1187$i$ & 0.7605$+$1.1053$i$ & 3.8401 & 0.8983$-$1.5598$i$ & $-5.5301$\\
   \hline 1.3354$+$0.1293$i$ & & 0.2671 & 3.0386 & $-4.9531$\\
   \hline 1.2557& 0.7742 & 1.2367 & 6.5979 & $-3.6350$\\
   \hline 1.3052$-$0.3104$i$ &  & 0.2949 & 1.7620$-$0.3678$i$ & $-3.3771$\\
   \hline 0.9287$+$0.4576$i$ & 1.5595$+$0.7685$i$ &5.1709& 1.4374$-$1.0835$i$ & $-2.0590$\\
   \hline 1.1841$+$0.6309$i$ & & 3.5520  & 0.2464 & $-1.4819$\\
   \hline 0.0000 & 0.0000 & 0.0000 & 0.0000 & $0.0000^1$\\
 \hline 0.0000& &0.0000 & 0.5377 & $0.0000^2$\\
 \hline & & 0.1207& 0.6181 & $0.0000^3$\\
 \hline
\end{tabular}
\caption{The numerical solutions of BAEs (\ref{new-BE-1}). The superscripts $\scriptstyle{1,\,2,\,3}$ are the symbols we added to distinguish the degenerate eigenvalue.}\label{new-table-1}
\end{center}
\end{table}

\begin{table}[htp]
\begin{center}
\small
\begin{tabular}{|c|c|c|}
\hline\rule{0pt}{13pt}
   $\overline{\mu}_1$ & $\overline{\mu}_2 $ &$E_{\overline L}$ \\[4pt]
 \hline & & $0.0000^1$\\
 \hline 0.5377& & $0.0000^2$\\
  \hline 0.1207& 0.6181 & $0.0000^3$\\
  \hline
\end{tabular}
\caption{The numerical solutions of BAEs (\ref{new-BE-2}).}\label{new-table-2}
\end{center}
\end{table}

\begin{table}[htp]
\begin{center}
\small
\begin{tabular}{|c|c|c|c|c|c|}
\hline \rule{0pt}{13pt}
   $\widetilde\lambda_1$ & $\widetilde\lambda_2 $ & $\widetilde\lambda_2 $ & $\widetilde\mu_1$ &$\widetilde\mu_2$ & $E_{\overline L}$\\[4pt]
 \hline 1.1572$-$0.6788$i$ & & & & & $-5.5301$\\
 \hline 3.3759 & 1.1572$-$0.6788$i$ &~~~~~~& 2.8732 &~~~~~~ & $-4.9531$\\
 \hline 2.3221 & 1.1572$-$0.6788$i$ & & 1.7395$+$0.4628$i$ & & $-3.3771$\\
 \hline 3.3759& & & & & $-3.6390$\\
 \hline 2.3221& & & & & $-2.0590$\\
 \hline 3.3759 & 2.3221 &  & 3.6054 & & $-1.4819$\\
  \hline
\end{tabular}
\caption{The numerical solutions of BAEs (\ref{new-BE-3}).}\label{new-table-3}
\end{center}
\end{table}

From Table \ref{new-table-1} we find that BAEs (\ref{new-BE-1})
have several special solutions: $\l_j\!=\!0$, $j=1,\ldots,N\!-\!M$ and $\mu_k\!=\!0,\,k=1,\ldots,n,\,1\!\leq\!n\!\leq\!N$. In these cases, the $T$-$Q$ relation (\ref{new-TQ-1}) reduces to the $T$-$Q$ relation (\ref{new-TQ-2}).

\begin{figure}[htp]
  \centering
  \includegraphics[width=11.cm]{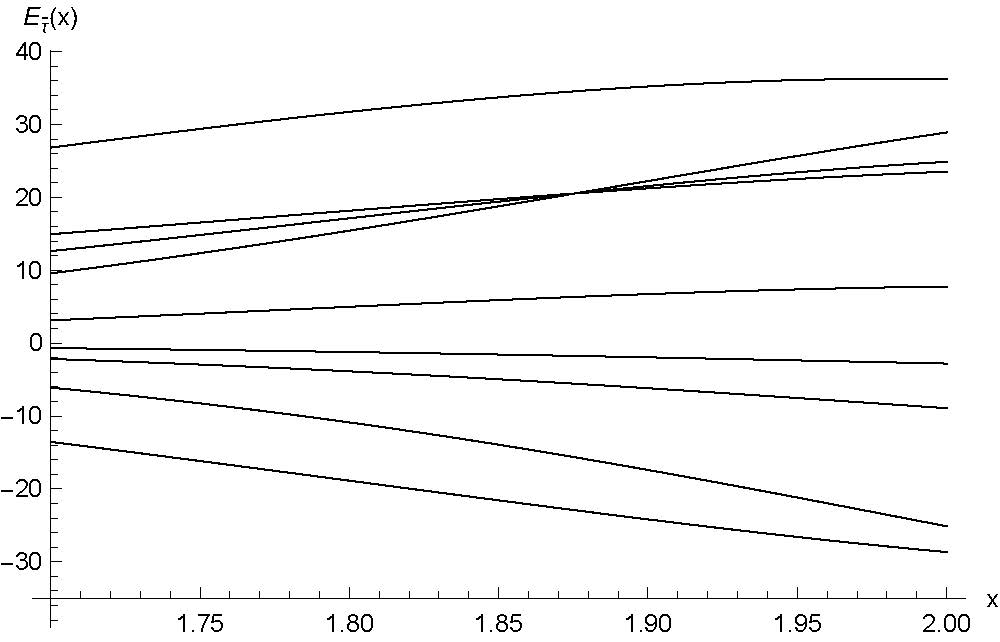}\\

  \caption{The eigenvalue curves of transfer matrix $\overline\tau(x)$ obtained from the exact diagonalization of $\overline\tau(x)$.} \label{figure of new transfer matrix}
\end{figure}

\begin{figure}[htp]
  \centering
  \includegraphics[width=13cm]{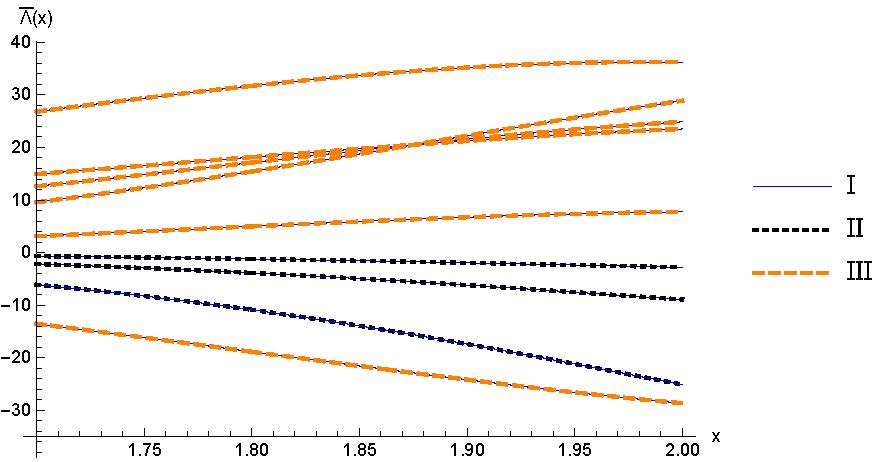}\\
  \caption{The eigenvalue curves of transfer matrix parameterized by the $T$-$Q$ realtions
 (\ref{new-TQ-1}), (\ref{new-TQ-2}) and (\ref{new-TQ-3}) which are labelled by ``I'', ``II'' and ``III'' respectively.}\label{figure of new TQ}
\end{figure}

For $N\!=\!2$ case, the eigenvalues of transfer matrix $\overline\tau(x)$ can be obtained by the exact diagonalization of $\overline\tau(x)$ or the $T$-$Q$ relations which are shown in Fig. \ref{figure of new transfer matrix} and Fig. \ref{figure of new TQ} respectively. The curves calculated from $T$-$Q$ relations and the nested BAEs are exactly the same as those obtained from the exact diagonalization of the transfer matrix $\overline\tau(x)$.

\subsection{The degenerate case $M=0$}
 When $M\!=\!0$ the system degenerates into a one-species ASEP with open boundaries. Let $M\!=\!0$, the $T$-$Q$ relations (\ref{new-TQ-2}) and (\ref{new-TQ-3}) can then be combined together with the following identity
\be
\overline\Lambda(x)=\frac{q^3-x^2}{q^2-x^2}\Lambda^{(2)}(x)+q^{-N}x^4y_2(x)f(q/x)z(qx).\label{relation-two-tau}
\ee
The function $\Lambda^{(2)}(x)$ is the eigenvalue of the transfer matrix $\tau^{(2)}(x)$ that corresponds to the single-species ASEP with open boundary conditions defined by
\be
\tau^{(2)}(x)={\rm tr}_0\left\{\mathcal{K}_0^{+}(x)\mathcal R_{0N}(x/\theta_N)\cdots \mathcal R_{01}(x/\theta_1)\mathcal{K}_0^{-}(x)\mathcal R_{10}(x\theta_1)\cdots \mathcal R_{N0}(x\theta_N)\right\},\label{nested-tau}
\ee
where
\be
&&\mathcal R(x)=\left(
             \begin{array}{cccc}
               a(x) & 0 & 0 & 0 \\
               0 & b^-(x) & c^+(x) & 0 \\
               0 & c^-(x) & b^+(x) & 0 \\
               0 & 0 & 0 & a(x) \\
             \end{array}
           \right),\label{nested-R}\\[4pt]
&&{\mathcal K}^{-}(x)=\left(
                \begin{array}{cc}
                  q(\g\hspace{-0.05cm}-\hspace{-0.05cm}\a)x^2\hspace{-0.05cm}+\hspace{-0.05cm}\eta_1x  & q\g(x^2\hspace{-0.05cm}-\hspace{-0.05cm}1) \\
                  q\a(x^2\hspace{-0.05cm}-\hspace{-0.05cm}1)  & \eta_1x\hspace{-0.05cm}+\hspace{-0.05cm}q(\g\hspace{-0.05cm}-\hspace{-0.05cm}\a) \\
                \end{array}
              \right),\label{nested-K-}\\[4pt]
&&\mathcal{K}^{+}(x)=\left(
                \begin{array}{cc}
                  \eta_2x\hspace{-0.05cm}+\hspace{-0.05cm}q^2(\d\hspace{-0.05cm}-\hspace{-0.05cm}\b)  & \b(x^2\hspace{-0.05cm}-\hspace{-0.05cm}q^2) \\
                  q\d(x^2\hspace{-0.05cm}-\hspace{-0.05cm}q^2)  & q(\d\hspace{-0.05cm}-\hspace{-0.05cm}\b)x^2\hspace{-0.05cm}+\hspace{-0.05cm}q\eta_2x \\
                \end{array}
              \right).\label{nested-K+}
\ee

When we adopt the integrable boundary conditions (\ref{K--1}) and (\ref{K+-1}), the total number of particle labelled by ``0'' is no longer conserved. However, this open 2-ASEP will also degenerate into the open ASEP defined by (\ref{nested-tau}) if there are no particles labelled by ``0''. Although we do not have an immediately intuitive meaning of the integer $M$ in Section \ref{T-Q relation}, we can also combine the $T$-$Q$ relation (\ref{TQ-1-1}) and (\ref{TQ-2}) together with an identity similar to (\ref{relation-two-tau}) when $M\!=\!0$,
\be
\Lambda(x)=\frac{q^3-x^2}{q^2-x^2}\Lambda^{(2)}(x)+q^{-N+4}y_2(x)f(x/q)z(qx).\label{new-relation-two-tau}
\ee

For the open 2-ASEP that we study in this paper, the eigenvalues of transfer matrix can be parameterized by several homogeneous $T$-$Q$ relations, which means that we can diagonalize the $K^+$-matrix and triangularize the $K^-$-matrix simultaneously.

\section{Integrable multi-species ASEP}\label{integrable-mASEP}\setcounter{equation}{0}

\paragraph{Higher rank $K$-matrices}
The $K$-matrices in (\ref{K--1}) and (\ref{NK-}) can be generalised to arbitrary multi-species ASEPs in the following manner,
\be
&&K^{(m=2r,-)}(x)^i_{j}=\left\{
                       \begin{array}{ll}
                         q(\g-\a)x^2+\eta_1 x, & ~~~~~1\hspace{-0.08cm}\leq \hspace{-0.08cm}i\hspace{-0.08cm}=\hspace{-0.08cm}j\hspace{-0.08cm}\leq \hspace{-0.08cm}r, \\
    \eta_1 x+q(\g-\a), & ~~~~~ r\hspace{-0.08cm}+\hspace{-0.08cm}2\hspace{-0.08cm}\leq \hspace{-0.08cm}i\hspace{-0.08cm}=\hspace{-0.08cm}j\hspace{-0.08cm}\leq \hspace{-0.08cm}2r\hspace{-0.08cm}+\hspace{-0.08cm}1,\\
q\g (x^2-1), & ~~~~~ 1\hspace{-0.08cm}\leq \hspace{-0.08cm}i\hspace{-0.08cm}\leq \hspace{-0.08cm}r,\quad \hspace{-0.08cm}j\hspace{-0.08cm}=\hspace{-0.08cm}2r\hspace{-0.08cm}+\hspace{-0.08cm}2\hspace{-0.08cm}
-\hspace{-0.08cm}i,\\
q\a(x^2-1), & ~~~~~ 1\hspace{-0.08cm}\leq\hspace{-0.08cm} j\hspace{-0.08cm}\leq\hspace{-0.08cm} r,\quad \hspace{-0.08cm}i\hspace{-0.08cm}=\hspace{-0.08cm}2r\hspace{-0.08cm}+\hspace{-0.08cm}2\hspace{-0.08cm}
-\hspace{-0.08cm}j,\\
q\g x^2+\eta_1 x-q\a, & ~~~~~ i=\hspace{-0.08cm}j\hspace{-0.08cm}=\hspace{-0.08cm}r\hspace{-0.08cm}+\hspace{-0.08cm}1,\\
0, &  ~~~~~\hbox{otherwise},
                       \end{array}
                     \right.\label{ms-K-1}\\[4pt]
&&K^{(m=2r-1,-)}(x)^i_{j}=\left\{
                       \begin{array}{ll}
                         q(\g-\a)x^2+\eta_1 x, &  ~~~~~1\hspace{-0.08cm}\leq \hspace{-0.08cm}i\hspace{-0.08cm}=\hspace{-0.08cm}j\hspace{-0.08cm}\leq \hspace{-0.08cm}r, \\
    \eta_1 x+q(\g-\a), & ~~~~~ r\hspace{-0.08cm}+\hspace{-0.08cm}1\hspace{-0.08cm}\leq \hspace{-0.08cm}i\hspace{-0.08cm}=\hspace{-0.08cm}j\hspace{-0.08cm}\leq \hspace{-0.08cm}2r,\\
q\g (x^2-1), &  ~~~~~1\hspace{-0.08cm}\leq \hspace{-0.08cm}i\hspace{-0.08cm}\leq \hspace{-0.08cm}r,\quad \hspace{-0.08cm}j\hspace{-0.08cm}=\hspace{-0.08cm}2r\hspace{-0.08cm}+\hspace{-0.08cm}1\hspace{-0.08cm}
-\hspace{-0.08cm}i,\\
q\a (x^2-1), & ~~~~~ 1\hspace{-0.08cm}\leq \hspace{-0.08cm}j\hspace{-0.08cm}\leq \hspace{-0.08cm}r,\quad \hspace{-0.08cm}i\hspace{-0.08cm}=\hspace{-0.08cm}2r\hspace{-0.08cm}+\hspace{-0.08cm}1\hspace{-0.08cm}
-\hspace{-0.08cm}j,\\
0, &  ~~~~~\hbox{otherwise},
                       \end{array}
                     \right.\label{ms-K-2}\\[4pt]
&&K^{(m,-)}(x)^i_j=\left\{
                      \begin{array}{ll}
                        q(\gamma-\alpha)x^2+\eta_1 x, &  ~~~~~i\hspace{-0.08cm}=\hspace{-0.08cm}j
                        \hspace{-0.08cm}=\hspace{-0.08cm}1,  \\
                        \eta_1 x+q(\gamma-\alpha), &  ~~~~~i\hspace{-0.08cm}=j\hspace{-0.08cm}
                        =\hspace{-0.08cm}m\hspace{-0.08cm}+\hspace{-0.08cm}1, \\
                        q\gamma(x^2-1), & ~~~~~ i\hspace{-0.08cm}=\hspace{-0.08cm}1,\quad \hspace{-0.08cm}2\hspace{-0.08cm}\leq\hspace{-0.08cm} j\hspace{-0.08cm}\leq\hspace{-0.08cm}m\hspace{-0.08cm}+\hspace{-0.08cm}1, \\
                        q\alpha(x^2-1), & ~~~~~ i\hspace{-0.08cm}=\hspace{-0.08cm}r,\quad\hspace{-0.08cm} 1\hspace{-0.08cm}\leq j\hspace{-0.08cm}\leq\hspace{-0.08cm} m, \\
                        -q\alpha x^2+\eta_1 x+q\gamma, & ~~~~~ 2\hspace{-0.08cm}\leq\hspace{-0.08cm} i\hspace{-0.08cm}=\hspace{-0.08cm}j\hspace{-0.08cm}\leq\hspace{-0.08cm} m, \\
                        0, &  ~~~~~\hbox{otherwise},
                      \end{array}
                    \right.\label{ms-K-3}
\ee
where $r$ is an integer and the parameter $\eta_1$ is given by (\ref{def-eta}).
The method proposed in this paper can be directly generalized to m-ASEP by constructing a set of fused transfer matrices $\{\tau(x),\ldots,\tau_{m+1}(x)\}$ when we adopt the $K$-matrices in (\ref{ms-K-1})-(\ref{ms-K-3}).

\paragraph{Other $K$-matrices for m-ASEP}
The other two $K$-matrices for 2-ASEP in \cite{Crampe} can also be generalised to m-ASEP as follow
\be
&&K^{(m,-)}(x)^i_{j}=\left\{
                      \begin{array}{ll}
                        q(\g-\a)x^3+\eta_1 x^2, & ~~~~~ i\hspace{-0.08cm}=\hspace{-0.08cm}j\hspace{-0.08cm}=\hspace{-0.08cm}1, \\
q\g x(x^2-1),& ~~~~~ i\hspace{-0.08cm}=\hspace{-0.08cm}1,\quad \hspace{-0.08cm}2\hspace{-0.08cm}\leq \hspace{-0.08cm}j\hspace{-0.08cm}\leq\hspace{-0.08cm} m\hspace{-0.08cm}+\hspace{-0.08cm}1,\\
q\a x(x^2-1),& ~~~~~ i\hspace{-0.08cm}=\hspace{-0.08cm}2,\quad \hspace{-0.08cm}j\hspace{-0.08cm}=\hspace{-0.08cm}1,\\
\eta_1 x^2+q(\g-\a)x, & ~~~~~i\hspace{-0.08cm}=\hspace{-0.08cm}j\hspace{-0.08cm}=\hspace{-0.08cm}2,\\
q\bar\a(x^2-1),&  ~~~~~i\hspace{-0.08cm}=\hspace{-0.08cm}2,\quad \hspace{-0.08cm}3\hspace{-0.08cm}\leq \hspace{-0.08cm}j\hspace{-0.08cm}\leq\hspace{-0.08cm}m\hspace{-0.08cm}+\hspace{-0.08cm}1,\qquad\\
(q-q\a x/\bar\a)(\g x+\bar\a),&  ~~~~~3\hspace{-0.08cm}\leq\hspace{-0.08cm} i\hspace{-0.08cm}=\hspace{-0.08cm}j\hspace{-0.08cm}\leq \hspace{-0.08cm}m\hspace{-0.08cm}+\hspace{-0.08cm}1,\\
0,& ~~~~~\hbox{otherwise},
                      \end{array}
                    \right.\label{ms-K-4}
                    \ee
                    \be
&&K^{(m,-)}(x)^i_{j}=\left\{
                      \begin{array}{ll}
                        \eta_1 x+q(\g-\a), & ~~~~~i\hspace{-0.08cm}=\hspace{-0.08cm}j\hspace{-0.08cm}=\hspace{-0.08cm}
                        m\hspace{-0.08cm}+\hspace{-0.08cm}1, \\
q\a (x^2-1),&~~~~~i=m\hspace{-0.08cm}+\hspace{-0.08cm}1,\quad 1\hspace{-0.08cm}\leq \hspace{-0.08cm}j\hspace{-0.08cm}\leq \hspace{-0.08cm}m,\\
q\g(x^2-1),& ~~~~~i\hspace{-0.08cm}=\hspace{-0.08cm}m,\quad \hspace{-0.08cm}j\hspace{-0.08cm}=\hspace{-0.08cm}m\hspace{-0.08cm}+\hspace{-0.08cm}1,\\
q(\g-\a)x^2+\eta_1 x, &~~~~~i\hspace{-0.08cm}=\hspace{-0.08cm}j\hspace{-0.08cm}=\hspace{-0.08cm}m,\\
q\bar\g x(x^2-1),& ~~~~~i\hspace{-0.08cm}=\hspace{-0.08cm}m,\quad \hspace{-0.08cm}1\hspace{-0.08cm}\leq \hspace{-0.08cm}j\hspace{-0.08cm}\leq \hspace{-0.08cm}m\hspace{-0.08cm}-\hspace{-0.08cm}1,\qquad\\
(q\g/\bar\g x-qx^2)(\bar\g x+\a),& ~~~~~1\hspace{-0.08cm}\leq\hspace{-0.08cm} i\hspace{-0.08cm}=\hspace{-0.08cm}j\hspace{-0.08cm}\leq \hspace{-0.08cm}m\hspace{-0.08cm}-\hspace{-0.08cm}1,\\
0,&~~~~~\hbox{otherwise},
                      \end{array}
                    \right.\label{ms-K-5}
\ee
where $\a\!=\!\bar\a+\frac{\bar\a(q-1)}{q(\g+\bar\a)}$ and $\g\!=\!\bar\g+\frac{\bar\g(1-q)}{q(\a+\bar\g)}$.
The parameters $\bar \a$ and $\bar\g$ are defined in terms of $\a$, $\g$ and the bulk transition rate in (\ref{ms-K-4}) and (\ref{ms-K-5}). If we adopt the $K$-matrices in (\ref{ms-K-4}) and (\ref{ms-K-5}), the eigenvalue of corresponding transfer matrix is a polynomial of higher degree, as a function of the spectral parameter $x$, compared to (\ref{ms-K-1})--(\ref{ms-K-3}).
Some additional operator identities are therefore needed to construct the corresponding $T$-$Q$ relation.

\paragraph{The construction of integrable m-ASEP}
We now show that there exist recursive rules for the construction of integrable m-ASEP. The basic boundary Markovian matrix is
\be
L_b^{(1)}=\left(
            \begin{array}{cc}
              -q\alpha & q\gamma \\
              q\alpha & -q\gamma \\
            \end{array}
          \right),
\ee
which corresponds to one-species ASEP. Suppose $L^{(m)}_{b,1}$ and $L^{(m)}_{b,2}$ are integrable boundary Markovian matrices for m-ASEP where $L^{(1)}_{b,1}\equiv L^{(1)}_{b,2}\equiv L^{(1)}_{b}$, then the integrable structures for higher rank ASEP can be constructed as \cite{cantini2016,crampe2016integrable}
\be
\begin{aligned}
&~~~~~~L_{b,1}^{(m+1)}=
\left(
  \begin{array}{c|c}
     L_{b,1}^{(m)} & \\[2pt]
     \hline
     & 0
  \end{array}
\right),\\[4pt]
&~~~~~~L_{b,k}^{(m+2)}=
\left(
  \begin{array}{c|c}
     L_{b,k}^{(m)} & \\[2pt]
     \hline \rule{0pt}{13pt}
     & L_{b}^{(1)}
  \end{array}
\right),\quad\qquad k=1,2,\\[4pt]
&~~~~~~L_{b,k}^{(m+1)}=
\left(
  \begin{array}{c|c}
    ~~~~~~~~ L_{b,k}^{(m)} ~~~~~~~~
    & \begin{array}{c}
        q\g' \\
        q\a' \\
        0\\
        \vdots\\
        0
      \end{array}
    \\[2pt]
    \hline
     & -q\sigma'
  \end{array}
\right),\qquad k=1,2,\\[4pt]
&~~~~~~L_{b,2}^{(m+1)}=
\left(
  \begin{array}{c|c}
    ~~~~~~~~ L_{b,2}^{(m)} ~~~~~~~~
    & \begin{array}{c}
        0\\
        \vdots\\
        0\\
        q\g \\
        q\a \\
      \end{array}
    \\[2pt]
    \hline
     & -q\sigma
  \end{array}
\right)~{\mbox {with $L_{b,2}^{(m)}=\left(
                                      \begin{array}{c|c}
                                        * &  \\
                                        \hline \rule{0pt}{14pt}& L_{b}^{(1)} \\
                                      \end{array}
                                    \right),
$}}
\end{aligned}
\ee
where all suppressed matrix elements are zero, $\sigma=\a+\g$, $\sigma'=\a'+\g'$ and $\a'$, $\gamma'$ satisfy one of the following set of constraints
\be
\begin{aligned}
& \a'=\a,\quad \gamma=\g'+\frac{\g'(1-q)}{q(\a+\g')},\\
& \a=\a'+\frac{\bar\a'(q-1)}{q(\g+\bar\a')},\quad \g'=\g.
\end{aligned}
\ee

\section*{Conclusion}
Using the nested off-diagonal Bethe Ansatz method, we find the Bethe ansatz solution for the spectrum of two integrable two-species ASEPs with open boundary conditions. We can use several homogeneous $T$-$Q$ relations to parameterize the eigenvalues of the transfer matrix. An interesting result are the identities (\ref{relation-two-tau}) and (\ref{new-relation-two-tau}), which directly relate the $T$-$Q$ relations for ASEP and 2-ASEP with open boundaries. A further work would be to analyse the other two boundary Markovian matrices given in \cite{Crampe}.

The method employed in this paper can be generalized to m-ASEP and other high rank integrable systems with open boundaries.
A set of commutative fused transfer matrices $\{\tau(x),\,\tau_2(x),\,\ldots,\,\tau_{m+1}(x)\}$ should be constructed. The $T$-$Q$ relations then can be found from similar considerations as in this paper. As a stochastic process, the boundary conditions of m-ASEP are constrained compared to their quantum spin chain analogous. Therefore we expect there also exist more than one homogeneous $T$-$Q$ relations for multi-species ASEP with open boundaries beyond rank one.

There is a gauge transformation between open ASEP and the spin-$1/2$ quantum XXZ
model with boundary terms \cite{essler1996}. Such a relation is still less known for the higher rank cases. The $K$-matrix (\ref{K--1}) has a different structure than the generic $K$-matrices for the trigonometric $SU(3)$ quantum spin chain \cite{li2016}. As a consequence, a new relation should be established
between open 2-ASEP and trigonometric $SU(3)$ spin chain.
We hope our result is helpful to answer this question.

The homogeneous $T$-$Q$ relations and the corresponding BAEs also allow one to further analyse the spectrum in the thermodynamic limit and to study physical properties of the system \cite{jan2005,jan2006}.

\section*{Acknowledge}

This contribution is in memory of our dear friend Vladimir Rittenberg. We gratefully acknowledge support from the Australian Research Council Centre of
Excellence for Mathematical and Statistical Frontiers (ACEMS). It is a
pleasure to thank Junpeng Cao, Alexandr Garbali, Kun Hao, Yupeng Wang, Michael Wheeler and Wen-Li Yang for discussing the calculations and results in this paper. We warmly thank Matthieu Vanicat for his advice on the $K$-matrices during the 2018 MATRIX program \textit{Non-equilibrium systems and special functions}.

\appendix
\section{Proof of asymptotic behavior}\label{A.1}\setcounter{equation}{0}
Rewrite the one-row monodromy matrices in matrix form
\be
\begin{aligned}
&T(x)=\left(
       \begin{array}{ccc}
         A_1(x) & B_1(x) & B_2(x) \\
         C_1(x) & D_1(x) & B_3(x) \\
         C_2(x) & C_3(x) & D_2(x) \\
       \end{array}
     \right),\\
&\widehat T(x)=\left(
       \begin{array}{ccc}
         \overline A_1(x) & \overline B_1(x) & \overline B_2(x) \\
         \overline C_1(x) & \overline D_1(x) & \overline B_3(x) \\
         \overline C_2(x) & \overline C_3(x) & \overline D_2(x) \\
       \end{array}
     \right).
\end{aligned}
\ee
The expression of the matrix $G$ is
\be
G=\a\b q\left(\mathbf{D}_2\mathbf{C}_1-\mathbf{D}_1\mathbf{C}_1+q^N\mathbf{B}_3\mathbf{D}_1^{-1}
-q^{N+1}\mathbf{B}_3\mathbf{D}_2^{-1}+\mathbf{B}_3\mathbf{C}_1\right),\label{def-G}
\ee
where the matrices in (\ref{def-G}) are defined by
\be
\begin{aligned}
&\mathbf{D}_1=\lim_{x\rightarrow\infty}\frac{D_1(x)}{(-x)^N}=d_1^{(1)}\ldots d_N^{(1)},\\
&\mathbf{D}_2=\lim_{x\rightarrow\infty}\frac{D_2(x)}{(-x)^N}=d_1^{(2)}\ldots d_N^{(2)},\\
&\mathbf{B}_3=\lim_{x\rightarrow\infty}\frac{B_3(x)}{(-x)^N}=\sum_{n=1}^Nd_1^{(1)}\cdots d_{n-1}^{(1)}b_n^{(3)}d_{n+1}^{(2)}\cdots d_N^{2)},\\
&\mathbf{C}_1=\lim_{x\rightarrow\infty}\frac{\overline C_1(x)}{(-x)^N}
=\sum_{n=1}^N a_N^{(2)}\cdots a_{n+1}^{(2)}c_n^{(1)}a_{n-1}^{(1)}\cdots a_1^{(1)}.
\end{aligned}
\ee
Here, $d^{(1)}={\rm diag}\{q,\,1,\,1\}$, $d^{(2)}={\rm diag}\{q,\,q,\,1\}$, $a^{(1)}={\rm diag}\{1,\,q,\,q\}$, $a^{(2)}={\rm diag}\{1,\,1,\,q\}$, $c^{(1)}=(1-q)E^{(1,2)}$, $b^{(3)}=(1-q)E^{(3,2)}$
and $E^{(i,j)}$ denotes the elementary $3\times3$ matrix with a single non-zero
entry $1$ at position $(i,j)$.
Obviously $G$ is a $3^N\times 3^N$ matrix in $\mathbf{V}_1\otimes \mathbf{V}_2\ldots\otimes \mathbf{V}_N$, we can find that only four types of matrix elements are nonzero:
\be
    &&G^{\,k_1,\ldots,1,\ldots,k_N}_{\,k_1,\ldots,2,\ldots,k_N},
    \quad G^{\,k_1,\ldots,3,\ldots,k_N}_{\,k_1,\ldots,2,\ldots,k_N},\quad
    G^{\,k_1,\ldots,1,\ldots,3,\ldots,k_N}_{\,k_1,\ldots,2,\ldots,2,\ldots,k_N},\quad G^{\,k_1,\ldots,3,\ldots,1,\ldots,k_N}_{\,k_1,\ldots,2,\ldots,2,\ldots,k_N}.
\ee
The positions of these non-zero elements imply that
$G$ has no contribution to the diagonalization of matrix $t_1$. With a similar procedure, we can prove that $\overline G$ doesn't contribute to the diagonalization of matrix $t_2$.

Due to the fact that $W$ is a diagonal matrix
we can rewrite $t_1$ as
\be
t_1=\underbrace{q^2\gamma\delta+q^{N+1}\a\b {W}+q^{N+2}\a\b {W}^{-1}}_{\mbox{a diagonal matrix}}~~~~~~~~~+~~~ \underbrace{~~~~~~G~~~~~~}_{\mbox{an aditional term}}.
\ee
The eigenvalues of $W$ are $\left\{q^M|M\!=\!0,1,\ldots,N\right\}$, thus we can easily diagonalize the matrices $t_1$ and $t_2$.

\section{Operators identities at special points}\label{A.2}\setcounter{equation}{0}
For convenience, define the following parameters
\begin{equation}\label{parameters}
\begin{aligned}
&e^{\pm}_1=\left(q\g\pm\eta_1-q\a\right),\\
&e_2^{\pm}=\left(q^3\d\pm q^{\frac52}\eta_2-q^4\b\right),\\
&e_3^{\pm}=\frac{q^3-1}{q-1}\left(q\d\pm\eta_2-q\b\right),\\
&e_4^{\pm}=\frac{1-q^3}{q^2-q^3}\left(q^3\g\pm q^{\frac52}\eta_1-q^4\a\right),\\
&e_5^{\pm}=(q^4-1)(q^3-1)(q\g\pm q\eta_1-q^3\a)(q^2\g\pm q\eta_1-q^2\a),\\
&e_6^{\pm}=(q^4-1)(q^3-1)(q\d\pm q^{\frac12}\eta_2-q^2\b)(q^2\d\pm q^{\frac12}\eta_2-q\b).
\end{aligned}
\end{equation}
The properties of $R$-matrix (\ref{initial-condition})-(\ref{crossing-unitary-condition}) and $K^{\pm}$-matrices (\ref{property-K})
allow us to calculate the values of transfer matrices $\tau(x)$ and $\tau_2(x)$ at some special points
\be
\begin{aligned}\label{SP}
&\tau(\pm1)=e_1^{\pm}z(\pm 1)\,{\rm tr}\{K^{+}(\pm1)\}\times \mathbb{I}=e_1^{\pm}\,e_3^{\pm}z(\pm 1)\times \mathbb{I},\\[4pt]
&\tau(\pm q^{\frac32})=q^{-N}e_2^{\pm}z(\pm q^{\frac52})\,{\rm tr}\left\{UK^-(\pm q^{\frac32})\right\}\times \mathbb{I}=q^{-N}e_2^{\pm}\,e_4^{\pm}z(\pm q^{\frac52})\times \mathbb{I},\\[4pt]
&\tau_2(\pm 1)=e_1^{\pm}\rho_2(q)z(\pm 1)\,{\rm tr}\left\{K^+(\pm 1)\right\}\tau(\pm q)=e_1^{\pm}\,e_3^{\pm}\rho_2(q)z(\pm 1)\tau(\pm q),\\[4pt]
&\tau_2(\pm q^{\frac12})=q^{-N}e_2^{\pm}\rho_2(q^2)z(\pm q^{\frac52})\,{\rm tr}\left\{K^-(\pm q^{\frac32})U\right\}\tau(\pm q^{\frac12})\\[4pt]
&\hspace{1.43cm}=q^{-N}e_2^{\pm}\,e_4^{\pm}\rho_2(q^2)z(\pm q^{\frac52})\tau(\pm q^{\frac12}),\\[4pt]
&\tau_2(\pm q)=q^{-2N}h_2(\pm q^2)z(\pm q^2)z(\pm q^3)\,
{\rm tr}_{12}\left\{{K^-_{12}}(\pm q)R_{12}(1)U_1U_2P_{21}^-\right\}\times \mathbb{I}\\[4pt]
&\hspace{1.20cm}=q^{-2N}h_2(\pm q^2)e_5^{\pm}z(\pm q^2)z(\pm q^3)\times \mathbb{I},\\[4pt]
&\tau_2(\pm q^{-\frac12})=h_1(\pm q^{\frac12})z(\pm q^{\frac12})z(\pm q^{-\frac12})
\,{\rm tr}_{12}\left\{K^+_{12}(\pm q^{\frac12})R_{21}(1)P_{12}^-)\right\}\times \mathbb{I}\\[4pt]
&\hspace{1.65cm}=h_1(\pm q^{\frac12})\,e_6^{\pm}z(\pm q^{\frac12})z(\pm q^{-\frac12})\times \mathbb{I},\\[4pt]
&\tau_2(\pm q^{-1})=\tau_2(\pm q^{\frac32})=0,
\end{aligned}
\ee
where function $z(x)$ is defined by (\ref{function-z}). The fused transfer matrices $\tau(x)$ and $\tau_2(x)$ share the same eigenstate, so the corresponding eigenvalues $\L(x)$ and $\L_2(x)$ satisfy the same functional relations
\be
\begin{aligned}\label{functional-SP}
&\Lambda(\pm1)=k_1^{\pm}\,k_3^{\pm}z(\pm 1),\\[4pt]
&\Lambda(\pm q^{\frac32})=q^{-N}k_2^{\pm}\,k_4^{\pm}z(\pm q^{\frac52}),\\[4pt]
&\Lambda_2(\pm 1)=k_1^{\pm}\,k_3^{\pm}\rho_2(q)z(\pm 1)\Lambda(\pm q),\\[4pt]
&\Lambda_2(\pm q^{\frac12})=q^{-N}k_2^{\pm}\,k_4^{\pm}\rho_2(q^2)z(\pm q^{\frac52})\Lambda(\pm q^{\frac12}),\\[4pt]
&\Lambda_2(\pm q)=q^{-2N}h_2(\pm q^2)\,k_5^{\pm}z(\pm q^2)z(\pm q^3),\\[4pt]
&\Lambda_2(\pm q^{-\frac12})=h_1(\pm q^{\frac12})\,k_6^{\pm}z(\pm q^{\frac12})z(\pm q^{-\frac12}),\\[4pt]
&\Lambda_2(\pm q^{-1})=\Lambda_2(\pm q^{\frac32})=0.
\end{aligned}
\ee

\end{document}